\documentclass[a4paper,fleqn]{cas-dc}

\usepackage[numbers]{natbib}
\usepackage{multirow}
\usepackage{color,soul}
\usepackage{graphicx}
\usepackage[utf8]{inputenc} %
\usepackage{graphicx, caption} %
\usepackage{threeparttable}%
\usepackage{lipsum}
\usepackage{caption}
\def\tsc#1{\csdef{#1}{\textsc{\lowercase{#1}}\xspace}}
\tsc{WGM}
\tsc{QE}
\tsc{EP}
\tsc{PMS}
\tsc{BEC}
\tsc{DE}

\begin{document}
\let\WriteBookmarks\relax
\def\floatpagepagefraction{1}
\def\textpagefraction{.001}
\shorttitle{A Deep Learning Approach for Parallel Imaging and Compressed Sensing MRI Reconstruction}
\shortauthors{F. Sadik et~al.}

\title [mode = title]{A Deep Learning Approach for Parallel Imaging and Compressed Sensing MRI Reconstruction}
\tnotemark[1]

\tnotetext[1]{This research did not receive any specific grant from funding agencies in the public, commercial, or non-profit sectors.}

\author[1]{Farhan Sadik}[orcid=0000-0001-5637-6101]

\author[1]{Md. Kamrul Hasan}[orcid=0000-0002-4816-2725]
\ead{khasan@eee.buet.ac.bd}
\cormark[1]

\address{Department of Electrical and Electronic Engineering,\\ Bangladesh University of Engineering and Technology,\\Dhaka 1205, Bangladesh}
\cortext[cor1]{Corresponding author}
\begin{abstract}
Parallel imaging accelerates MRI data acquisition by acquiring additional sensitivity information with an array of receiver coils, resulting in fewer phase encoding steps. Because of fewer data requirements than parallel imaging, compressed sensing magnetic resonance imaging (CS-MRI) has gained popularity in the field of medical imaging. Parallel imaging and compressed sensing (CS) both reduce the amount of data captured in the \textit{k}-space, which speeds up traditional MRI acquisition. As acquisition time is inversely proportional to sample count, forming an image from reduced \textit{k}-space samples results in faster acquisition but with aliasing artifacts. For de-aliasing the reconstructed image, this paper proposes a novel Generative Adversarial Network (GAN) called RECGAN-GR that is supervised with multi-modal losses. In comparison to existing GAN networks, our proposed method introduces a novel generator network, RemU-Net, which is integrated with dual-domain loss functions such as weighted magnitude and phase loss functions, as well as parallel imaging-based loss, GRAPPA consistency loss. As refinement learning, a \textit{k}-space correction block is proposed to make the GAN network self-resistant to generating unnecessary data, which speeds up the reconstruction process. Comprehensive results show that the proposed RECGAN-GR not only improves the PSNR by 4 dB over GAN-based methods but also by 2 dB over conventional state-of-the-art CNN methods available in the literature for single-coil data. The proposed work significantly improves image quality for low-retained data, resulting in five to ten times faster acquisition.

\end{abstract}



\maketitle
\section{Introduction}
\label{sec1}
Magnetic resonance imaging provides high-quality
images to attain comprehensive information about the vasculature of
a region of living tissue and microstructural features 
non-invasively. High contrast anatomical details can be brought into
focus leading to detect tissue injuries, stroke, and dementia
\cite{Revett2011} without exposing patients to 
radiation. 

Nonetheless, prolonged acquisition time due to the sampling process in the \textit{k}-space domain leads to difficulty in maintaining constant posture and causes severe motion artifacts. Therefore, to lessen patients' discomfort, accelerated MR image reconstruction with high resolution is of great necessity.

The MRI raw data samples are collected in the Fourier domain, known as \textit{k}-space and the inversion speed of \textit{k}-space to the image domain is limited by physiological and hardware constraints.
 Faster acquisition requires the violation of Nyquist-Shannon sampling theorem which results in an aliased image. Numerous techniques have emerged over the past years for de-aliasing the image and reducing the acquisition time while keeping a diagnostically suitable resolution. Hardware constrained parallel MRI (pMRI) \cite{Hamilton2017} works by acquiring a reduced amount of \textit{k}-space data with an array of multiple receiver coils.
 Several parallel imaging algorithms are used to reconstruct images which include SENSE \cite{Pruessmann1999,Sajal2019}-type reconstruction, where additional coil sensitivity information is used to eliminate the effect of aliasing in the image domain and, GRAPPA \cite{Griswold2002}-type reconstruction strategies that form the missing \textit{k}-space lines from known lines to perform the reconstruction process in the \textit{k}-space domain. A combination of both methods such as SPACE-RIP \cite{Kyriakos2000}, SPIRiT \cite{Murphy2012} provide flexibility to reconstruct \textit{k}-space lines from arbitrary sampled data along with additional image priors. However, SENSE-type strategies require accurate sensitivity maps and GRAPPA-type reconstructions require approximations which hinder the image quality. Another evolving strategy is known as compressed sensing MRI \cite{Lustig2008} which requires undersampling the data at a much lower rate than the Nyquist-Shannon theorem in some sparsifying transform domains. Dictionary learning and transfer learning have been in practice to reconstruct the image from aggressively undersampled data \cite{Ravishankar2011,Feng2013}. However, the clinical practice of CS-MRI is limited by manually adjustable parameters. 

Lately, deep learning methods have shown great potential for reconstructing missing lines from \textit{k}-space or as a post-processor to improve the quality of the reconstructed image. These methods can be abbreviated into four classes. Single domain flat CNN based methods, dual-domain CNN-based methods, flat unrolled iterative CNN and, Generative Adversarial Network-based algorithms.  
 Single domain flat CNN based methods require images to be obtained from the zero-filling reconstruction or GRAPPA \textit{k}-space reconstruction and inverse Fourier transform before feeding it to a Convolutional Neural Network (CNN). Jin $et\hspace{1mm} al.$ \cite{Jin2017} proposed to utilize U-Net architecture as an autoencoder with residual skip connections to enhance the zero-filled reconstructed image. Zeng $et\hspace{1mm} al.$ \cite{Zeng2019} proposed a very deep CNN which consists of several sub-networks performed in the image domain. Each sub-network consists of one multi-level feature extraction block and a data consistency (DC) layer. The feature extraction block generates the intermediate image and the DC block updates the image according to the raw image domain data. Transform domain flat CNN models exclusively operate in the frequency domain or \textit{k}-space domain. Missing lines are estimated using deep learning models and the image is reconstructed from the final output image of the network by applying inverse Fourier transform \cite{Han2020}. The quantitative results are poor as these methods do not fully exploit the inherent properties of the \textit{k}-space. 
 
 Dual-domain CNN-based methods utilize a cascade of two or more CNN networks and DC modules in between. Schlemper $et\hspace{1mm} al.$ \cite{Schlemper2018} proposed a cascade of two CNNs for de-aliasing, in parallel with DC layers to solve the CS-MRI optimization problem with a co-ordinate descent type algorithm until convergence. The DC layer imposes a constraint on the selectivity of known \textit{k}-space data. In short, the unknown samples of the undersampled \textit{k}-space are replaced by the output of the CNN and the known samples are expressed as a linear combination of the known measurement and the prediction of the CNN weighted by some noise. Eo $et\hspace{1mm} al.$ \cite{Eo2018} used two separate CNN networks, one in the \textit{k}-space domain and another in the spatial domain. Souza $et\hspace{1mm} al.$ \cite{Souza2020} utilized two U-Nets in a similar fashion. Both the models are trained end-to-end integrating interleaved DC layers between dual-domain networks. Wang $et\hspace{1mm} al.$ \cite{Wang2019} proposed two sequential CNN, one in the \textit{k}-space domain and another in the spatial domain linked by DC layers. Moreover, multi supervised loss function strategy is utilized, i.e., spatial loss and \textit{k}-space loss coupled by some weights along with existing mean-square-loss (MSE) loss. Ran $et\hspace{1mm} al.$ \cite{Ran2021} introduced two parallel branches of CNNs operating simultaneously on both domains and interaction between them occurs through data consistency layers in both domains. The dual-domain fusion layer combines the result from the parallel branches to acquire the final reconstructed image. The results of this method are still not up to the mark compared to other methods but leaves a room for improvement as this kind of method has not been broadly studied so far. Flat unrolled iterative CNN such as Yan $et\hspace{1mm} al.$ \cite{NIPS2016_1679091c}, divides the inverse problem of CS-MRI into three sub-problems that correspond to reconstruction, denoising, and multiplier update operation which is known as the Alternating Direction Method of Multipliers (ADMM) operation. The ADMM blocks are integrated into a data flow graph that consists of several stages where each stage corresponds to a specific ADMM operation and the network is optimized through cascaded deep neural networks for parameter learning. In GAN-based algorithms, a generalized ill-posed inverse problem is solved which is similar to the image super-resolution \cite{Dong2016} problem. The zero-filled reconstructed image is passed through the generator network and a mini-max game is played between the generator and the discriminator until there is an optimal adversarial loss where the reconstructed image closely resembles the ideal image provided in the training process. Quan $et\hspace{1mm} al.$ \cite{Quan2018} introduced a U-Net shaped generator optimized with frequency loss and image loss along with the adversarial loss in a cyclic fashion. Yang $et\hspace{1mm} al.$ \cite{Yang2018} incorporated VGG-based perceptual loss along with the spatial and frequency domain losses. The non-trainable VGG works like a human eye scoring the perceptual quality of the reconstructed image. The weighted sum of these losses is minimized and binary cross-entropy is used as the adversarial loss. Jiang $et\hspace{1mm} al.$ \cite{Jiang2019} performed some tuning on \cite{Yang2018} which includes cascading three U-Nets in the generator network instead of one. Wasserstein distance is utilized \cite{Vallender1974} as the adversarial loss to stabilize the GAN network. One of the major disadvantages of U-Net shaped generator architectures is the heavy information loss in the encoding decoding process which hinders the reconstruction performance \cite{Ran2021}. Due to fewer constraints, GAN-based networks converge before reaching to an optimal reconstruction performance. The coordination between the GAN-based architectures with pMRI strategies is not fully exploited yet. 

In this paper, we propose a novel GAN-based parallel imaging and compressed sensing MRI reconstruction scheme, namely RECGAN-GR. A novel generator network, RemU-Net is introduced to restrain the loss of details along the encoder-decoder path of the generator. In addition, a parallel imaging-based loss function, e.g., GRAPPA consistency loss is proposed along with the \textit{k}-space magnitude and phase correction losses with different weights to preserve subtle anatomical details at the generator.
Furthermore, a \textit{k}-space correction block is proposed to limit the generator from producing unnecessary lines, resulting in the reconstruction of the missing lines only. This improves the convergence profile and ensures faster reconstruction.
\section{Methodology}\label{sec2}
In this section, the prerequisites of the proposed RECGAN-GR method e.g., parallel MRI, CS-MRI, and GANs are briefly explained.
\subsection{Classic Parallel MRI}\label{subsec2}
Parallel MRI (pMRI) uses spatial sensitivity information built in an array of multiple receiver surface coils for image acquisition and acquires a fraction of phase encoding steps to reduce scan time. Each coil is sensitive to a specific region of the anatomical structure described by the weight ``coil sensitivity''. Parallel imaging-based techniques such as SENSE \cite{Pruessmann1999} use the pixel-position wise multiplication of the original object image with the coil sensitivity maps to get the coil images and sum-of-squares technique to obtain the desired image from a reduced FOV image.  
 The consequence of the multiplication of coil sensitivity with the full FOV image in \textit{k}-space domain is the spreading of \textit{k}-space coil information across neighboring \textit{k}-space points. As \textit{k}-space points can be identified from neighboring \textit{k}-space points, \textit{k}-space-based reconstruction approaches exploit this idea to reconstruct missing lines from the acquired lines.
Unlike image domain reconstructions, \textit{k}-space reconstruction utilizes additional calibration lines in the center region of the \textit{k}-space, or auto-calibration signals (ACS), which are collected along with the reduced data acquisition. One of the common methods of \textit{k}-space reconstruction is generalized autocalibrating partially parallel acquisitions
(GRAPPA) \cite{Griswold2002}. In this method, data is acquired from multiple lines from all coils to fit a single line in a coil. This process is continued for all the coils for generating coil images and the sum-of-squares (SoS) technique is used to combine these coil images into a single image. The GRAPPA reconstruction formalism can be represented as
\begin{equation}
{
 \vec{K}^{(u)}=\mathcal{G}*\vec{K}
}
\end{equation}
where $\vec{K}^{(u)}$ is the unknown \textit{k}-space line which is a linear combination of GRAPPA weighted neighboring acquired \textit{k}-space lines ($\vec{K}$) from all coils. It may be seen as a complex-valued convolution in \textit{k}-space from $N_{c}$ channels to $N_{c}$ channels, where $N_{c}$ is the number of coils. The missing \textit{k}-space lines can be reconstructed convolving acquired \textit{k}-space lines with GRAPPA weight vector $\mathcal{G}$. The fully sampled data (i.e., ACS) which corresponds to the low spatial frequencies in the \textit{k}-space are used to estimate the GRAPPA weights $\mathcal{G}$. We can estimate $\mathcal{G}$ by solving the following optimization problem:
\begin{equation}\label{4}
{
 \hat{\mathcal{G}}=\min_{\mathcal{G}}\|\vec {K^{(u)^{\prime}}}-\mathcal{G}*{\vec{K^{\prime}}}\|^{2}
}
\end{equation}
where $\vec{K^{\prime}}$ is the observed and
$\vec{K^{(u)^{\prime}}}$ is the unobserved \textit{k}-space lines in the ACS, respectively.
 
\subsection{Compressed Sensing MRI}\label{subsec3}
 MR images can be sparsely represented in some transform domain which makes MRI a natural fit for compressed sensing. Let $\bold{x}$ $\epsilon$  $\mathbb{C}^{N_{x}N_{y}}$ denotes an MR image and $\bold{y}_{f}$ $\epsilon$ $\mathbb{C}^{N_{x}N_{y}}$ is the fully sampled \textit{k}-space representation of $\bold{x}$, where $N_{x}$ $\times$ $N_{y}$ is the image size. The corresponding undersampled \textit{k}-space data is $\bold{y}$ $\epsilon$ $\mathbb{C}^{N_{x}N_{y}}$. The zero-filled reconstructed image from the undersampled \textit{k}-space data $\bold{y}$ is represented by $\bold{x}_{u}$ $\epsilon$ $\mathbb{C}^{N_{x}N_{y}}$. The problem can be formulated as
\begin{equation}\label{5}
{
 \bold{y}=\bold{F}_{u}\bold{x} + \boldsymbol{\varepsilon}
} 
\end{equation}
where $\bold{F}_{u}$ is known as Fourier encoding matrix and $\boldsymbol{\varepsilon}$ is the acquisition noise. $\bold{F}_{u}$ is equal to the Fourier transform matrix elementwise multiplied with the undersampled mask ($\bold{F}_{u} = \bold{F}\odot\bold{U}$). The inverse problem is ill-posed and it has many solutions even without the noise part. To encounter this problem, some prior estimate of $\bold{x}$ is required which can be expressed as the following optimization problem:
\begin{equation} \label{6}
{
\min_{\bold{x}} \frac{1}{2} \|\bold{F}_{u}\bold{x}-\bold{y}\|_{2}^{2} + \lambda \Re(\bold{x})
} 
\end{equation}
where $\frac{1}{2} \|\bold{F}_{u}\bold{x}-\bold{y}\|_{2}^{2}$ is the data fidelity term and  $\Re$ is the prior regularisation term that depends on $\bold{x}$. $\lambda$ $\geq$ 0 is a factor that determines the balance between the data fidelity term and the prior regularisation term. $\Re$($\bold{x}$) is generally an $l_{0}-$ or $l_{1}-$ norm in a sparsifying transform domain such as Fourier, Wavelet or Discrete Cosine Transform.

\subsection{DL Based CS-MRI}\label{subsec4}
 Deep learning based CS-MRI approaches \cite{Yang2018,Wang2019,Schlemper2018,Ran2021} incorporate an additional regularization term with Eqn. \ref{6} in order to find a unique solution to the reconstruction problem as in the following:
 \begin{equation} \label{7}
{
\min_{x} \frac{1}{2} \|\bold{F}_{u}\bold{x}-\bold{y}\|_{2}^{2} + \lambda \Re(\bold{x}) + \zeta \|\bold{x}-f_{cnn}(\bold{x}_{u} \lvert \boldsymbol{ \hat\Theta})\|_{2}^{2}
} 
\end{equation}
where  the input to the model is $\bold{x}_{u}$ and $ f_{cnn}(\bold{x}_{u} \lvert \boldsymbol{\hat\Theta})$ is the output of the CNN with parameter $\boldsymbol{\hat \Theta}$. $\zeta$ is a regularization parameter. The goal is to find an optimal $\boldsymbol{\hat \Theta}$ to reconstruct an image with minimal loss. In this work, the "integration model" \cite{7493320} approach is incorporated where the output of the CNN ($f_{cnn}(\bold{x}_{u} \lvert \boldsymbol{ \hat\Theta})$) is used as a reference image to add up as a regularization term along with the CS-MRI optimization function.

MRI data usually consists of two complex planes which are real plane and imaginary plane. Two approaches exist for deep learning methods to handle complex numbers: (1) real plane is embedded into the complex plane using an operator Re* : $\mathbb{R}^{N}$ $\mapsto$ $\mathbb{C}^{N}$ where Re*(\textbf{x}) = \textbf{x} + 0i, (2) both planes are treated separately. In our method, we use the real plane for the CNN input but for loss functions, both of the planes are utilized.
\subsection{Contributions of The Proposed Method}\label{subsec5}
The contributions of the proposed method are summarized below:
\begin{enumerate}
 
\item  A RemU-Net generator network consistent with the U-Net architecture is proposed.
 
\item \textit{K}-space loss functions are utilized with separate weights for magnitude and phase, illustrating the significance of phase reconstruction for better image quality.

\item A novel GRAPPA consistency loss function for both the image domain and \textit{k}-space domain is introduced.

\item For refinement learning, a \textit{k}-space correction block is introduced which forces the generator to construct only the missing lines in the frequency domain. 
\item 
Pixel-wise mean-square loss (MSE) and perceptual loss that use pre-trained weights from the VGG \cite{DBLP:journals/corr/SimonyanZ14a} (Visual Geometry Group) are coupled with the proposed loss functions to construct an end-to-end MRI reconstruction.

\end{enumerate}

 
\begin{figure*}[h]
    \centering
    \includegraphics[width=0.9\linewidth]{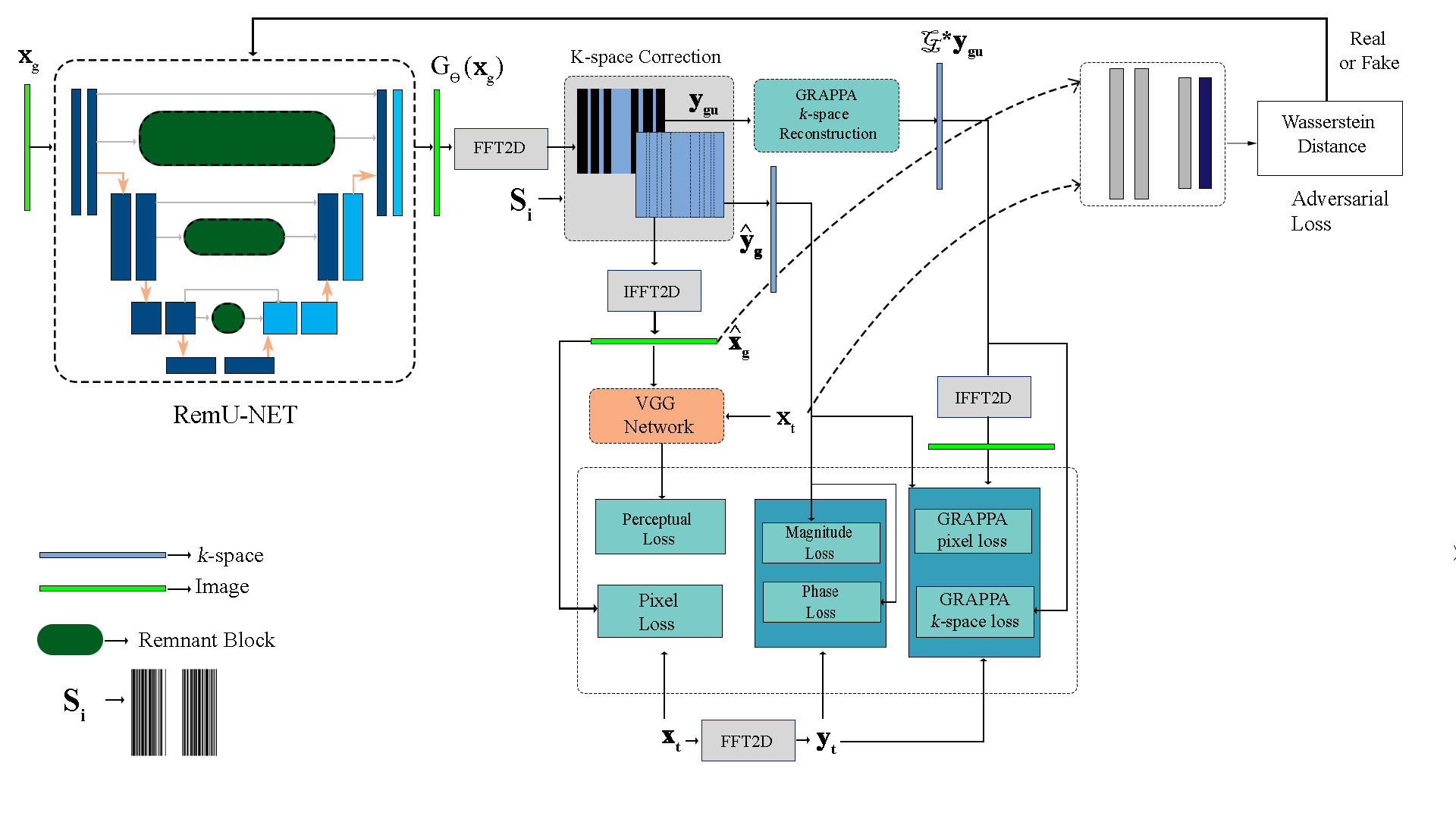}
    \caption{Illustration of the proposed RECGAN-GR architecture for MRI reconstruction.}
    \label{fig:RECGAN_GR_final}
\end{figure*}
The proposed scheme is shown in Fig. \ref{fig:RECGAN_GR_final}.

\section{The Proposed Pipeline}\label{sec4}
\subsection{The Proposed Generator Architecture}\label{subsec1}
In order to understand the proposed RemU-Net generator, the basic GAN architecture is discussed prior to describing this network.
\subsubsection{General GAN}\label{subsubsec1}
Two basic building blocks of generative adversarial network \cite{NIPS2014_5ca3e9b1} are a generator $G(\bold{z},\theta_G)$ and a discriminator $D(\bold{x},\theta_D)$. The generator outputs a fake data $G_{\theta_{G}}(\bold{z})$, where $\bold{z}$ is an input prior noise variable and the discriminator calculates the distance between the synthesized fake data $G_{\theta_{G}}(\bold{z})$ and the orginal data $\bold{x}$. Both networks play a mini-max game to find the optimal parameters $\theta_G$ and $\theta_D$, e.g., the generator outputs a fake data $(G_{\theta_{G}}(\bold{z}):\bold{z}\rightarrow\bold{x})$ and the discriminator ($D_{\theta_{D}}(\bold{x}):\bold{x}\rightarrow[0,1]$) tries to find the error between the real data and the synthetic data. The game ends when the discriminator cannot distinguish two variables which means the generated image closely resembles the original one. The training process can be formulated as
\begin{multline}\label{8}
\min_{\theta_{G}} \max_{\theta_{D}} \mathscr{L}(\theta_{G},\theta_{D})=\mathbb{E}_{x}\sim p_{r}(\bold{x})[\mathrm{log}  D_{\theta_{D}}(\bold{x})]+ \\
\mathbb{E}_{z}\sim p_{z}(\bold{z})[\mathrm{log}(1-  D_{\theta_{D}}(G_{\theta_{G}}(\bold{z})))] 
\end{multline}
where $p_{z}(\bold{z})$ is the latent probability distribution of the noise variable $\bold{z}$ and $p_{r}(\bold{x})$ is the distribution of the original data. Here, $\mathbb{E}$ means mathematical expectation. For CS-MRI the GAN model is initialized with some prior information also known as conditonal GAN \cite{Isola2017} that is

\begin{multline}\label{9}
\min_{\theta_{G}} \max_{\theta_{D}} \mathscr{L}_{cGAN}(\theta_{G},\theta_{D})=\mathbb{E}_{x_{t}}\sim p_{train}(\bold{x}_{t})\\
[\mathrm{log}  D_{\theta_{D}}(\bold{x}_{t})]+
\mathbb{E}_{x_{u}}\sim p_{G}(\bold{x}_{u})[\mathrm{log}(-  D_{\theta_{D}}(G_{\theta_{G}}(\bold{x}_{u})))] 
\end{multline}
where $\bold{x}_{u}$ is the input to the generator which is the zero-filled aliased image. The output of the generator is the de-aliased image $\bold{\hat x}_{u}$ and the discriminator tries to differentiate between the generated output and the target image $\bold{x}_{t}$. 
The optimal discriminator for the generator distribution $p_{G}(\bold{x}_{u})$ is found when the generator distribution resembles the original distribution that is $p_{train}(\bold{x}_{t})=p_{G}(\bold{x}_{u})$. The optimal discriminator is given by
\begin{equation}\label{11}
{
 \theta_{D}^{*}(\bold{x})=\frac{p_{train}(\bold{x}_{t})}{p_{train}(\bold{x}_{t})+p_{G}(\bold{x}_{u})}
}
\end{equation}
The optimal discriminator reduces the minimax game into a minimization over the generator only and is equal to 
\begin{equation}\label{12}
{
 \min_{\theta_{G}} \mathscr{L}(\theta^{*}_{D},\theta_{G})=JSD(p_{train}\parallel p_{G})-\mathrm{log}(2)
}
\end{equation}
where JSD stands for Jensen-Shannon divergence \cite{DBLP:journals/corr/ArjovskyB17} which is zero when two distributions are equal and non-zero otherwise. Wasserstein distance is utilized in the proposed method as adversarial loss which reduces the vanishing gradient problem and improves the training stability. The distance between the training data $p_{train}(\bold{x}_{t})$ and the synthetic data $p_{G}(\bold{x}_{u})$ can be formulated as
\begin{equation}\label{13}
{
 W(p_{train},p_{G}) =\inf_{\gamma \sim \Pi (p_{train},p_{G})} \mathbb{E}_{\bold{x}_{t},\bold{\hat x}_{u}\sim \gamma} [\|\bold{x}_{t}-\bold{\hat x}_{u}\|] 
}
\end{equation}
\subsubsection{RemU-Net Generator Network}\label{subsubsec2}
The major motivation behind the proposed RemU-Net generator network is to overcome the unreliability of the skip connections in U-Net used in DAGAN \cite{Yang2018} for the preservation of semantic informations. 
In contrast to the nested, dense skip pathways in U-Net++ \cite{Zhou2018}, the proposed RemU-Net architecture uses the remnant block \cite{Rafi2020} as the sub-network. It is consistent with the traditional U-Net where the encoder and decoder branch consists of  8 convolution layers and 8 deconvolution layers, respectively, followed by batch normalization (BN) and leaky ReLU as the activation function.

\begin{figure*}[h]
    \centering
    \includegraphics[width=0.9\linewidth]{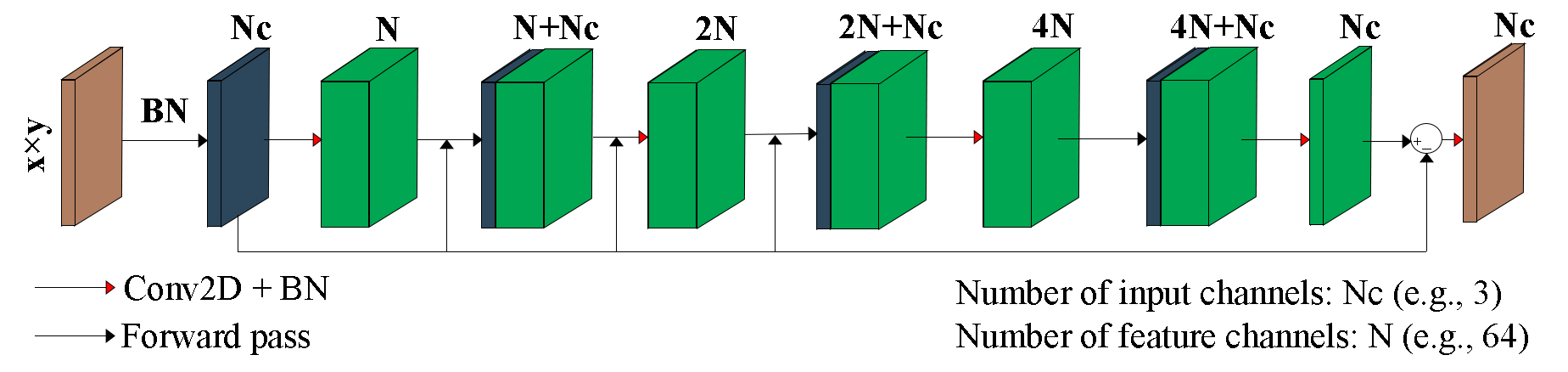}
    \caption{Modified remnant block for the RemU-Net generator network. The height and the width of the input is \textbf{x} and \textbf{y}, respectively.}
    \label{fig:Remnant}
\end{figure*}

 A modified remnant block preprocessor is proposed as a bridge between the encoder and decoder along with skip connections. It consists of 3 convolution layers with kernel size of 3 $\times$ 3 followed by BN. The input is batch normalized before passing into the remnant block. The feature channel is widened to 64, 128, 256, respectively in three successive convolution layers, and the batch normalized input is propagated through each layer to reduce information loss. Finally, subtracting the output of the remnant block from the input creates the residue image which contains the latent structure of the image regardless of the source. This residue is then concatenated with the decoder. The generator utilizes a hyperbolic tangent function as the output activation function. The block diagram of the modified remnant block is shown in Fig. \ref{fig:Remnant}. The input to the generator is a GRAPPA reconstructed image $\bold{x}_{g}$  from the undersampled \textit{k}-space data instead of the inverse Fourier transform of the zero filling reconstructed \textit{k}-space i.e., image $\bold{x}_{u}$. The generator continuously performs mapping G : $\bold{x}_{g}$ $\rightarrow$ $\bold{\hat x}_{g}$ and the discriminator is a classifier that calculates the error between $\bold{\hat x}_{g}$ and the target image $\bold{x}_{t}$. The discriminator architecture is the same as DAGAN \cite{Yang2018}.

\subsection{Proposed Multi-Strain Loss}\label{subsec3}
\begin{figure}[h]
    \centering
    \includegraphics[width=9 cm]{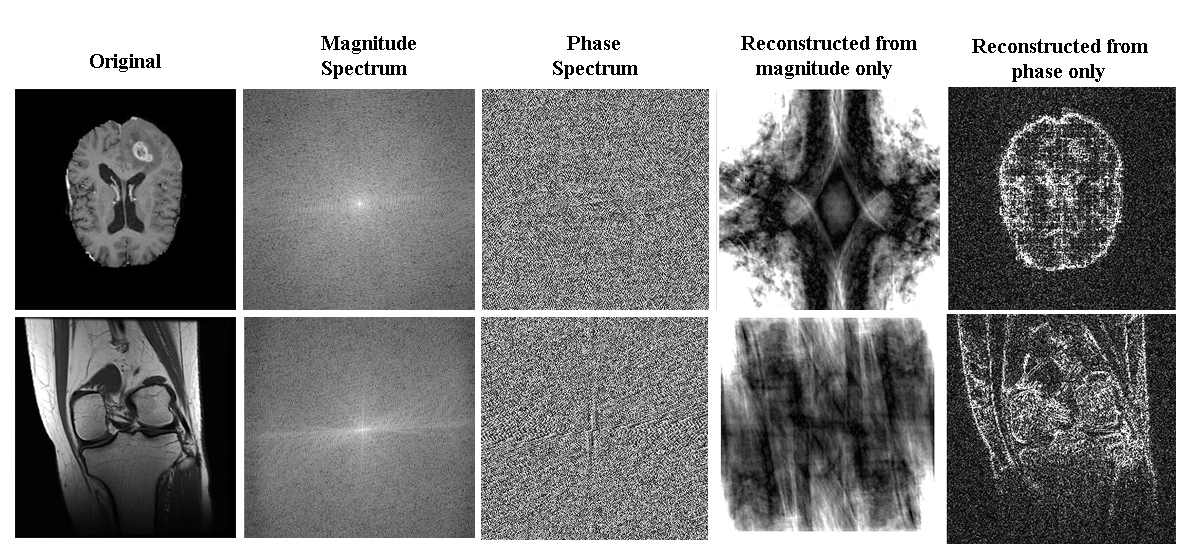}
    \caption{Examples of brain and knee MRI images and their magnitude and phase spectrum. Reconstructing the image with only magnitude and phase depicts the importance of preserving the phase information.}
    \label{fig:Phase_loss_figure}
\end{figure}

A multi-strain loss function is introduced that comprises losses from both spatial and \textit{k}-space domains. The loss comprises of five multi-disciplinary $l_{2}$ norms, namely, pixel-wise MSE, frequency domain loss in both magnitude and phase with different weighting factors, parallel imaging-based reconstruction loss, e.g., GRAPPA consistency loss in both spatial and \textit{k}-space domain, and perceptual loss. The idea of pixel-wise MSE, frequency domain loss, and VGG-based perceptual loss is derived from \cite{Yang2018,Jiang2019} and can be represented as:
\begin{equation}
{
  \min_{\theta_{G}}\mathscr{L}_{iMSE}(\theta_{G})=\frac{1}{2} \|\bold{\hat x}_{g}-\bold{x}_{t}\|_{2}^{2}
}
\end{equation}
\begin{equation}
{
  \min_{\theta_{G}}\mathscr{L}_{f_{M}MSE}(\theta_{G})=\frac{1}{2} \|\bold{\hat y}_{g}-\bold{y}_{t}\|_{2}^{2}
}
\end{equation}
\begin{equation} \label{perc}
{
  \min_{\theta_{G}}\mathscr{L}_{VGG}(\theta_{G})=\frac{1}{2} \| f_{vgg}(\bold{\hat x}_{g})-f_{vgg}(\bold{x}_{t})\|_{2}^{2}
}
\end{equation}
where $\bold{y}_{t}$ and $\bold{\hat y}_{g}$ are the frequency domain (magnitude) representation of the target image $\bold{x_{t}}$ and the de-aliased image $\bold{\hat x_{g}}$, respectively. We propose a phase spectrum loss function as edge features of an image are embedded in the phase spectrum \cite{OlivaresMercado2008}. As shown in Fig. \ref{fig:Phase_loss_figure}, the edge information lies in the phase spectrum and the magnitude spectrum does not contain any structural details.
The phase components can easily be obtained for both the target image and the synthetic image as

\begin{equation}
{
 \bold{y_{\theta_{t}}}=\bold{y_{t}} \cdot / \|\bold{y_{t}}\| 
}
\end{equation}
\begin{equation}
{
 \bold{\hat y_{\theta_{g}}}= \bold{\hat y_{g}} \cdot / \|\bold{\hat y_{g}}\| 
}
\end{equation}
where, $\bold{y_{\theta_{t}}}$ and $\bold{\hat y_{\theta_{g}}}$ are the phase spectrum of the Fourier transform of $\bold{y_{t}}$ and $\bold{\hat y_{g}}$ respectively. The proposed phase loss function is the $l_{2}$-norm of these two spectral components given by: 
\begin{equation}
{
  \min_{\theta_{G}}\mathscr{L}_{f_{\theta} MSE}(\theta_{G})=\frac{1}{2} \|\bold{\hat y}_{\theta_{g}} -  \bold{y}_{\theta_{t}} \|_{2}^{2}
}
\end{equation}
Parallel imaging-based reconstruction method GRAPPA reconstructs the unobserved \textit{k}-space lines $(\vec{K^{(u)}})$ from the observed ACS lines ($\vec{K}$) by proper estimation of a kernel ($\mathcal{G}$). GRAPPA reconstruction in \textit{k}-space is used for 
estimating the unobserved \textit{k}-space lines from the observed lines. The GRAPPA consistency loss function in \textit{k}-space domain calculates the $l_{2}$-norm between the reconstructed \textit{k}-space and the true \textit{k}-space. Likewise, the GRAPPA consistency loss function in the spatial domain takes the $l_{2}$-norm between the inverse Fourier transform of the reconstructed \textit{k}-space and the true image.
The input to the GRAPPA reconstruction block is the undersampled \textit{k}-space generated from the de-aliased synthetic image $(G_{\theta_{G}}(\bold{x_{g}}))$. For simulation, we incorporate sampling masks to undersample the image and take the center dominant regions as ACS lines. Let, $\mathscr{\bold{S}}_{i}$ be the sampling mask and $\bold{y_{g_{u}}}$ is the undersampled \textit{k}-space image given by
\begin{equation}
{
  \bold{y}_{g_{u}}=\mathscr{\bold{S}}_{i}(\mathscr{\bold{F}}(G_{\theta_{G}}(\bold{x_{g}}))) + \boldsymbol{\eta}
}
\end{equation}
where $\mathscr{\bold{F}}$ is the Fourier transform operator and $\boldsymbol{\eta}$ is the synthetic acquisition noise. The GRAPPA kernel $\mathcal{G}$ estimates the missing \textit{k}-space points as a linear combination of neighboring weighted points from different coils. Both single-coil and multi-coil data are used for simulation study in our work. The GRAPPA reconstructed \textit{k}-space is represented by $\mathcal{G}* \bold{y}_{g_{u}}$ which is interpreted as the convolution of the undersampled \textit{k}-space data with the GRAPPA kernel. The \textit{k}-space GRAPPA reconstruction loss functions can be represented by

 \begin{equation}
{
   \min_{\theta_{G}}\mathscr{L}_{GRAPPA_{k}}(\theta_{G})=\frac{1}{2} \|\mathcal{G}*\bold{y}_{g_{u}}-\bold{y}_{t}\|_{2}^{2}
}
\end{equation}
Furthermore, taking the $l_{2}-norm$ of the difference between GRAPPA reconstructed \textit{k}-space and ground truth images, we get the GRAPPA consistency loss in the spatial domain given by

 \begin{equation}
{
   \min_{\theta_{G}}\mathscr{L}_{GRAPPA_{s}}(\theta_{G})=\frac{1}{2} \|\mathscr{\mathrm{Re}(\bold{F}}^{-1}(\mathcal{G}*\bold{y}_{g_{u}}))-\bold{x}_{t}\|_{2}^{2}
}
\end{equation}

The adversarial loss \cite{NIPS2014_5ca3e9b1} of the generator is denoted as
 \begin{equation}
{
   \min_{\theta_{G}}\mathscr{L}_{GEN}(\theta_{G})=-\mathrm{log}(D_{\theta_{D}}(G_{\theta_{G}}(x_{g})))
}
\end{equation}
The total loss function is the sum of these multi-strain losses given by

\begin{multline}
   \mathscr{L}_{TOTAL}=\alpha\mathscr{L}_{iMSE}+ \beta \mathscr{L}_{f_{M}MSE} +\gamma\mathscr{L}_{f_{\theta}MSE} + \\ 
   \delta \mathscr{L}_{GRAPPA_{s}} + \zeta \mathscr{L}_{GRAPPA_{k}} + \kappa \mathscr{L}_{VGG} + \mathscr{L}_{GEN}
   \end{multline}
The pixel-wise MSE loss combined with the adversarial loss of the generator forms the fundamental part of the total loss which serves to solve the third optimization term in \ref{7}. However, pixel-wise MSE results in nonsmooth reconstruction \cite{Yang2018}, therefore, the frequency domain loss functions, particularly the phase loss component, aid in bringing out the edges. Furthermore, the GRAPPA loss function serves as a linear optimization to exploit the inherent sensitivity information in the \textit{k}-space domain and helps the generator to perform robust reconstruction regardless of the nature of the dataset. Additionally, perceptual similarity is ensured through the VGG-based perceptual loss.
\subsection{\textit{K}-space Correction Block}\label{subsec4}
Stable training of GAN is a tremendous task due to the exploding gradient problem which occurs when the discriminator gets overfitted and the generator receives no gradients. Ordinarily, the inverse Fourier transform of the undersampled zero filling reconstructed \textit{k}-space is concatenated with the output of the generator is used to tackle this problem \cite{Yang2018}. Although this concatenation somewhat reduces the heavy training task, there is a high probability of mixing original \textit{k}-space lines with the generated noisy lines. In order to overcome this problem, we propose a \textit{k}-space correction block to not only stabilize GAN but also force the generator to preserve the observed \textit{k}-space lines and generate only missing lines eliminating redundancies. The proposed \textit{k}-space block is formulated as 
\begin{equation}
   \hat y_{g}(i,j)=\begin{cases}
   \mathscr{F} (G_{\theta_{G}}(\bold{x}_{g}(i,j))), & \text{ $if (i,j) \not \in \Omega$}\\
    \mathscr{F}(\bold{x}_{g}(i,j)), & \text{$ if (i,j) \in \Omega$ }
  \end{cases}
\end{equation}
where $\Omega$ is the observed \textit{k}-space points and (\textit{i},\textit{j}) denotes the pixel coordinates of the image in the \textit{k}-space domain. The reconstructed image can be found in the spatial domain $\bold{\hat x_{g}}$ by applying inverse Fourier transform operator on $\bold{\hat y_{g}}$. 
\section{Experimental Design}\label{sec5}
\subsection{Datasets}\label{5:subsec1}
In this paper, we trained and tested our model using BRATS MICCAI brain tumor dataset 2019\footnote{https://www.med.upenn.edu/cbica/brats2019/registration.html} \cite{Menze2015} for single-coil brain MR images. In addition, we employed MRNet Knee MRI dataset\footnote{https://stanfordmlgroup.github.io/competitions/mrnet/} from Stanford ML Group for knee dataset. For single-coil brain MRI simulation, we randomly selected 8,695 2D images for training and 1,806 images for testing. The Knee MRI dataset consists of three different plane images that are coronal, axial, and sagittal. We mixed all
three types and total 7,801 images were taken for training and
1,523 images for testing to emphasis generalizability of the 
proposed method. The size of each image slice is 240 $\times$ 240.

12-channel multi-coil data from the Calgary Campinas\footnote{ https://sites.google.com/view/calgary-campinas-dataset/}  dataset \cite{SOUZA2018482} was utilized to confirm that our suggested model is compatible with real-world pMRI acquisition. 10,820 images—8,820 for training and 2000 for validation—and 50 test images—were taken from the dataset. In keeping with the practical MRI acquisition, Cartesian masks rather than random masks were used in order to get the undersampled \textit{k}-space in the phase encoding direction from the fully sampled multi-coil \textit{k}-space complex data. The size of each image slice for this dataset is 218 $\times$ 170. However, resize operation is performed before feeding the images into the network.

\subsection{Simulation Configuration}\label{5:subsec2}
\begin{figure}[h]
\centering
\includegraphics[width=0.998\linewidth]{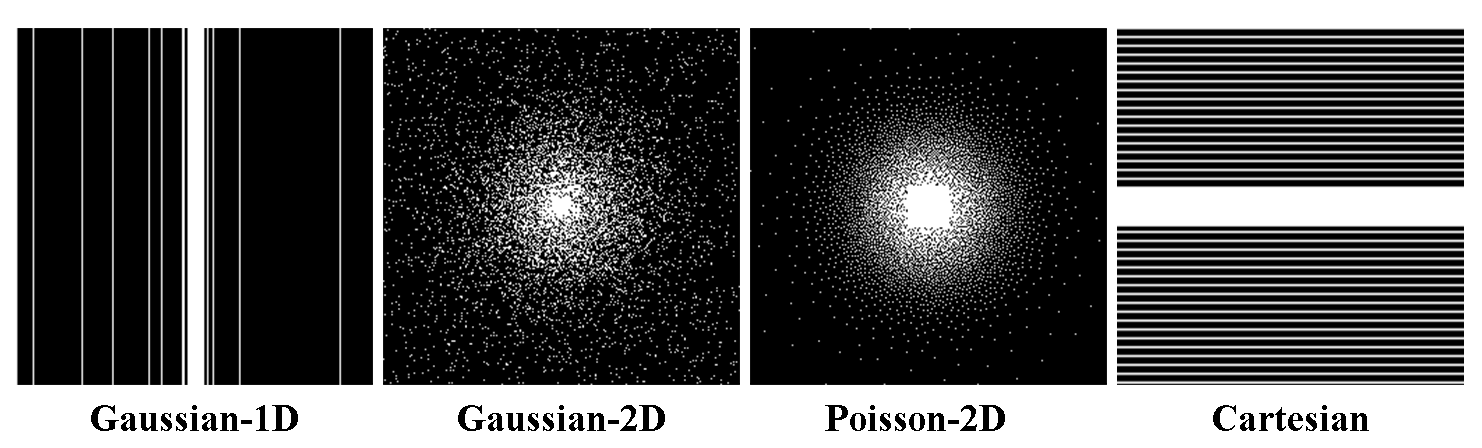}
\caption{Different sampling patterns used for simulation.}
\label{fig:sampling_patter}
\end{figure}

The models were implemented on a server configured on NVIDIA TESLA V-100 (32 GB), 384 GB RAM, and Windows Server 2019 operating system. The codes were implemented on Python and Tensorflow\footnote{https://www.tensorflow.org/} frameworks with CUDA and CUDNN support. To create the undersampled data from the fully sampled images, 1D pseudo-Gaussian sampling mask with 10\%, 20\%, 30\%, 40\%, and 50\% retained \textit{k}-space value are utilized that corresponds to 10$\times$, 5$\times$, 3.3$\times$, 2.5$\times$, 2$\times$ acceleration, respectively. For multi-coil data, Cartesian sampling masks are used where lines are skipped depending on the acceleration factor. The acceleration factor is defined as the ratio of \textit{k}-space lines needed for a fully sampled reconstruction to the collected \textit{k}-space lines in an accelerated acquisition \cite{deshmane2012parallel}. For example, acceleration factor, R means (R-1) number of subsequent lines on the \textit{k}-space are skipped during acquisition. To keep consistency with the GRAPPA-like reconstruction, a few lines in the central region of the \textit{k}-space are maintained as auto-calibration signal (ACS). The sampling is performed in the phase encoding direction. For each type of sampling technique, however, distinct training cycles are required. To compare with other state-of-the-art methods 2D Gaussian and 2D Poisson sampling masks (Fig. \ref{fig:sampling_patter}) were also utilized along with 1D masks. Random Gaussian noise is added with the masks in order to bear a resemblance to the practical data acquisition.
Adam optimizer with momentum 0.5 is used to train all the RECGAN-GR variations with parameters $\beta_{1}$ = 0.9, $\beta_{2}$ = 0.999, $\epsilon$ = 10$^{-8}$. The VGG-16 network pre-trained with the weights of the ImageNet \cite{ILSVRC15} and the output of the conv1, conv2, and conv4 layers are used to calculate the perceptual loss. Each variation was trained with the following hyperparameters: batch size = 1, initial learning rate = 0.0001. The weights associated with losses are: $\alpha$ = 16, $\beta$ = 0.2, $\gamma$ = 0.04, $\delta$ = 0.008, $\zeta$ = 0.00015 and $\kappa$ = 10$^{-3}$.
\subsection{Model Variations}\label{5:subsec3}
Five variations are tested in order to ensure the performance of the proposed RECGAN-GR. The variants are: (1) De-aliasing GAN with spatial loss and weighted phase loss (DAGSP), (2) DAGSP with weighted magnitude and phase loss (DAGSF), (3) DAGSF + GRAPPA consistency loss in both image and \textit{k}-space domain (DAGSF-GR),
(4) DAGSF-GR + RemU-Net (REMGSF-GR), and (5) REMGSF-GR + proposed \textit{k}-space correction block (RECGAN-GR). Each variation concludes the calculation of perceptual loss, e.g., VGG-based loss according to (\ref{perc}).
\subsection{Performance Evaluation}\label{5:subsec4}
For quantitative assessment of the reconstructed images, two commonly used metrics are measured known as peak signal to noise ratio (PSNR) and structural similarity index (SSIM) \cite{setiadi2021psnr}:
\begin{equation}
  \mbox{PSNR}=20\ \mathrm{log}_{10}\left( \frac{255}{ \sqrt{\frac{1}{MN} \sum_{i=0}^{M-1}\sum_ {j=0}^{N-1}(y(i,j)-x(i,j))^{2}}}\right)
\end{equation}
where $x$ and $y$ represent the fully sampled image and the reconstructed image, respectively and $i$ and $j$ denotes the coordinates of the pixels for image size \textit{M}$\times$\textit{N}. And
\begin{equation}
  \mbox{SSIM}= \frac{(2\mu_{x}\mu_{y}+C_{1})(2 \sigma_{xy}+C_{2})}{(\mu_{x}^{2}+\mu_{y}^{2}+C_{1})(\sigma_{x}^{2}+\sigma_{y}^{2}+C_{2})}
\end{equation}
where $\mu_{x}$ and $\mu_{y}$ are the means for the images $x$ and $y$, and that $\sigma_{x}$, $\sigma_{y}$ are the variances, respectively. $\sigma_{xy}$ is the co-variance between the two images. The constants $C_{1}$ and $C_{2}$ depends on the dynamic range of pixel values.
\subsection{Recent State-of-the-art Methods}\label{5:subsec5}
    The proposed RECGAN-GR is compared with five recent methods available in the literature which are DAGAN \cite{Yang2018}, DA-FWGAN \cite{Jiang2019}, RefineGAN \cite{Quan2018}, DC-CNN \cite{Schlemper2018}, W-Net IIII \cite{Souza2020}. RefineGAN, DAGAN, DA-FWGAN are GAN-based methods and the rest of the methods rely on convolutional neural networks with data consistency blocks. The methods under comparison were initialized with a zero filling solution of the reconstructed \textit{k}-space data by the authors. The proposed models, however, are initialized with GRAPPA reconstruction in the \textit{k}-space domain and later inverse Fourier transform is applied to feed the corresponding image to the neural networks. For fair comparison, we initialize all the aforesaid methods with GRAPPA reconstruction.   
\section{Results}\label{sec6}
\subsection{Ablation Study}\label{6:subsec1}
We segregated the testing to investigate the effectiveness of each step introduced in the proposed model and compare them to justify the selection of the final model. Tables \ref{tab:var_brain} and \ref{tab:knee_var}  summarize the statistical analysis of results on brain MRI and knee MRI data, respectively of the proposed model building blocks for the 1D Gaussian sampling pattern. 
\begin{table*}[h] \label{tab:var_brain}
\caption{Evaluation results of model variations for Brain MRI data with 1D Gaussian mask} 
\resizebox{18 cm}{1.5 cm}{
\centering 
\begin{tabular}{l ll ll ll ll ll} 
\hline 
Method  & \begin{tabular}{ c c  }
 \multicolumn{2}{c}{10\%}  \\ 
 \hline  PSNR & \hspace{0.8cm} SSIM  \\  
\end{tabular} &\begin{tabular}{ c c  }
 \multicolumn{2}{c}{20\%}  \\ 
 \hline  PSNR & \hspace{0.8cm} SSIM  \\  
\end{tabular}&\begin{tabular}{ c c  }
 \multicolumn{2}{c}{30\%}  \\ 
  \hline   PSNR & \hspace{0.8cm}  SSIM  \\  
\end{tabular} &\begin{tabular}{ c c  }
 \multicolumn{2}{c}{40\%}  \\ 
  \hline  PSNR & \hspace{0.8cm} SSIM  \\  
\end{tabular} &\begin{tabular}{ c c  }
 \multicolumn{2}{c}{50\%}  \\ 
  \hline  PSNR & \hspace{0.8cm}  SSIM  \\  
\end{tabular}
\\
\hline
 DAGSP  & \begin{tabular}{ c c }
 34.01 $\pm$ 2.01 & 0.924 
 \end{tabular}
   & \begin{tabular}{ c c   }
 38.52 $\pm$ 6.45 & 0.962 
 \end{tabular} & \begin{tabular}{ c c  }
 42.23 $\pm$ 4.02 & 0.992 
 \end{tabular} & \begin{tabular}{ c c  }
 43.36 $\pm$ 4.59 & 0.993 
 \end{tabular} & \begin{tabular}{ c c   }
 46.23 $\pm$ 3.89 & 0.995 
 \end{tabular}  \\
DAGSF & \begin{tabular}{ c c }
 34.23 $\pm$ 4.28 & 0.922 
 \end{tabular}
   & \begin{tabular}{ c c   }
 39.85 $\pm$ 4.20 & 0.966 
 \end{tabular} & \begin{tabular}{ c c  }
 43.03 $\pm$ 3.12 & 0.984 
 \end{tabular} & \begin{tabular}{ c c  }
 44.10 $\pm$ 4.26 & 0.988 
 \end{tabular} & \begin{tabular}{ c c   }
 46.15 $\pm$ 4.97 & 0.996 
 \end{tabular}  \\
 DAGSF-GR   & \begin{tabular}{ c c }
 34.51 $\pm$ 3.05 & 0.920 
 \end{tabular}
   & \begin{tabular}{ c c   }
 39.50 $\pm$ 4.23 & 0.931 
 \end{tabular} & \begin{tabular}{ c c  }
 45.02 $\pm$ 3.89 & 0.962 
 \end{tabular} & \begin{tabular}{ c c  }
 46.01 $\pm$ 3.42 & 0.987 
 \end{tabular} & \begin{tabular}{ c c   }
 48.53 $\pm$ 4.21 & 0.993 
 \end{tabular}  \\ 
REMGSF-GR  & \begin{tabular}{ c c }
 36.13 $\pm$ 2.51 & 0.967
 \end{tabular}
   & \begin{tabular}{ c c   }
 39.50 $\pm$ 2.25 & 0.978 
 \end{tabular} & \begin{tabular}{ c c  }
 45.65 $\pm$ 3.95 & 0.980 
 \end{tabular} & \begin{tabular}{ c c  }
 46.60 $\pm$ 4.29 & 0.984 
 \end{tabular} & \begin{tabular}{ c c   }
 48.52 $\pm$ 3.62 & 0.992 
 \end{tabular}  \\ 
 \hline
\textbf{RECGAN-GR}  & \begin{tabular}{ c c }
 \textbf{37.92 $\pm$ 3.41} & \textbf{0.972} 
 \end{tabular}
   & \begin{tabular}{ c c   }
 \textbf{40.23$\pm$ 3.85} & \textbf{0.982} 
 \end{tabular} & \begin{tabular}{ c c  }
 \textbf{46.20 $\pm$ 3.05} & \textbf{0.996} 
 \end{tabular} & \begin{tabular}{ c c  }
 \textbf{47.25 $\pm$ 4.35}  & \textbf{0.997}
 \end{tabular} & \begin{tabular}{ c c   }
 \textbf{49.87 $\pm$ 4.01} & \textbf{0.997} 
 \end{tabular}  \\ 
 \hline
\end{tabular}
}
\label{tab:var_brain}
\end{table*}
\begin{table*}[h] \label{tab:knee_var}
\caption{Evaluation results of model variations for Knee MRI data with 1D Gaussian mask} 
\resizebox{18cm}{1.5 cm}{
\centering 
\begin{tabular}{l ll ll ll ll ll} 
\hline 
Method  & \begin{tabular}{ c c }
 \multicolumn{2}{c}{10\%}  \\ 
 \hline  PSNR & \hspace{0.65cm} SSIM 
\end{tabular} &\begin{tabular}{ c c  }
 \multicolumn{2}{c}{20\%}  \\ 
 \hline  PSNR & \hspace{0.65cm} SSIM 
\end{tabular}&\begin{tabular}{ c c   }
 \multicolumn{2}{c}{30\%}  \\ 
  \hline   PSNR & \hspace{0.65cm}  SSIM  
\end{tabular} &\begin{tabular}{ c c   }
 \multicolumn{2}{c}{40\%}  \\ 
  \hline  PSNR & \hspace{0.65cm} SSIM  
\end{tabular} &\begin{tabular}{ c c }
 \multicolumn{2}{c}{50\%}  \\ 
  \hline  PSNR & \hspace{0.65cm}  SSIM  
\end{tabular}
\\
\hline
 DAGSP  & \begin{tabular}{ c c }
 33.79 $\pm$ 2.11 & 0.904  
 \end{tabular}
   & \begin{tabular}{ c c    }
38.12 $\pm$ 3.89 & 0.937 
 \end{tabular} & \begin{tabular}{ c c  }
 39.23 $\pm$ 4.52 & 0.960 
 \end{tabular} & \begin{tabular}{ c c  }
 40.05 $\pm$ 4.19 & 0.972 
 \end{tabular} & \begin{tabular}{ c c   }
 40.53 $\pm$ 3.09 & 0.980 
 \end{tabular}  \\
DAGSF & \begin{tabular}{ c c  }
   34.03 $\pm$ 1.78 & 0.917 
 \end{tabular}
   & \begin{tabular}{ c c  }
 38.50 $\pm$ 3.58 & 0.941 
 \end{tabular} & \begin{tabular}{ c c   }
 39.43 $\pm$ 3.25 & 0.965 
 \end{tabular} & \begin{tabular}{ c c  }
40.10 $\pm$ 2.16 & 0.976 
 \end{tabular} & \begin{tabular}{ c c   }
 41.28$\pm$ 3.19 & 0.982 
 \end{tabular}  \\
 DAGSF-GR & \begin{tabular}{ c c  }
 36.51 $\pm$ 3.05 & 0.920 
 \end{tabular}
   & \begin{tabular}{ c c    }
39.75 $\pm$ 4.23 & 0.947  
 \end{tabular} & \begin{tabular}{ c c   }
40.65 $\pm$ 3.89 & 0.962 
 \end{tabular} & \begin{tabular}{ c c   }
 41.56 $\pm$ 3.82 & 0.983 
 \end{tabular} & \begin{tabular}{ c c    }
42.23 $\pm$ 3.21 & 0.990 
 \end{tabular}  \\ 
REMGSF-GR  & \begin{tabular}{ c c  }
 38.53 $\pm$ 2.61 & 0.915 
 \end{tabular}
   & \begin{tabular}{ c c    }
 40.02 $\pm$ 2.25 & 0.978 
 \end{tabular} & \begin{tabular}{ c c   }
41.17 $\pm$ 1.95 & 0.980 
 \end{tabular} & \begin{tabular}{ c c }
42.60 $\pm$ 2.29 & 0.984 
 \end{tabular} & \begin{tabular}{ c c    }
 43.52 $\pm$ 1.62 & 0.992 
 \end{tabular}  \\ 
 \hline
\textbf{RECGAN-GR}  & \begin{tabular}{ c c  }
 \textbf{39.92 $\pm$ 2.43} & \textbf{0.921} 
 \end{tabular}
   & \begin{tabular}{ c c    }
 \textbf{40.05 $\pm$ 1.49} & \textbf{0.981} 
 \end{tabular} & \begin{tabular}{ c c   }
 \textbf{41.50 $\pm$ 1.67} & \textbf{0.992} 
 \end{tabular} & \begin{tabular}{ c c   }
 \textbf{42.89 $\pm$ 1.78}  & \textbf{0.994} 
 \end{tabular} & \begin{tabular}{ c c    }
 \textbf{44.62 $\pm$ 1.87} & \textbf{0.997} 
 \end{tabular}  \\ 
 \hline
\end{tabular}
}
\label{tab:knee_var}
\end{table*}
Implementation of both magnitude and phase in the frequency domain with different weights in DAGSF provides better quantitative results at the lower sampling rates than applying the phase loss only as in DAGSP for both brain and knee data. DAGSF-GR indicates the significance of GRAPPA reconstruction loss for better visual quality. The effect of GRAPPA consistency loss is significant for higher sampling rates in the brain data ($\approx$ 2 dB improvement) than lower sampling rates. However, for the knee MRI data, better reconstruction are also seen for lower sampling rates (e.g., $10\%$ sampling). For an acceleration factor of two (50\% retained raw data) the performance of GRAPPA is better than any other sampling rate and close to SENSE reconstruction \cite{Blaimer2004, Sajal2019} which is consistent with our finding. One of the major problems of U-Net shaped GAN for reconstruction is that during synthetic data generation, a significant amount of detail is lost in the encoder due to downsampling operation \cite{Ran2021}. REMGSF-GR incorporates RemU-Net, where remnant blocks are fused at different stages of the U-Net for better consistency between encoder and decoder to restore these missing details. The result is prominent even for the higher acceleration factor, e.g., 10\% retained data (Tables \ref{tab:var_brain} and \ref{tab:knee_var}). This method along with the proposed refinement learning, e.g., the \textit{k}-space correction (RECGAN-GR) gives the best reconstruction quality in terms of PSNR.
\begin{figure}[t!]
    \centering
    \includegraphics[width=0.9998\linewidth]{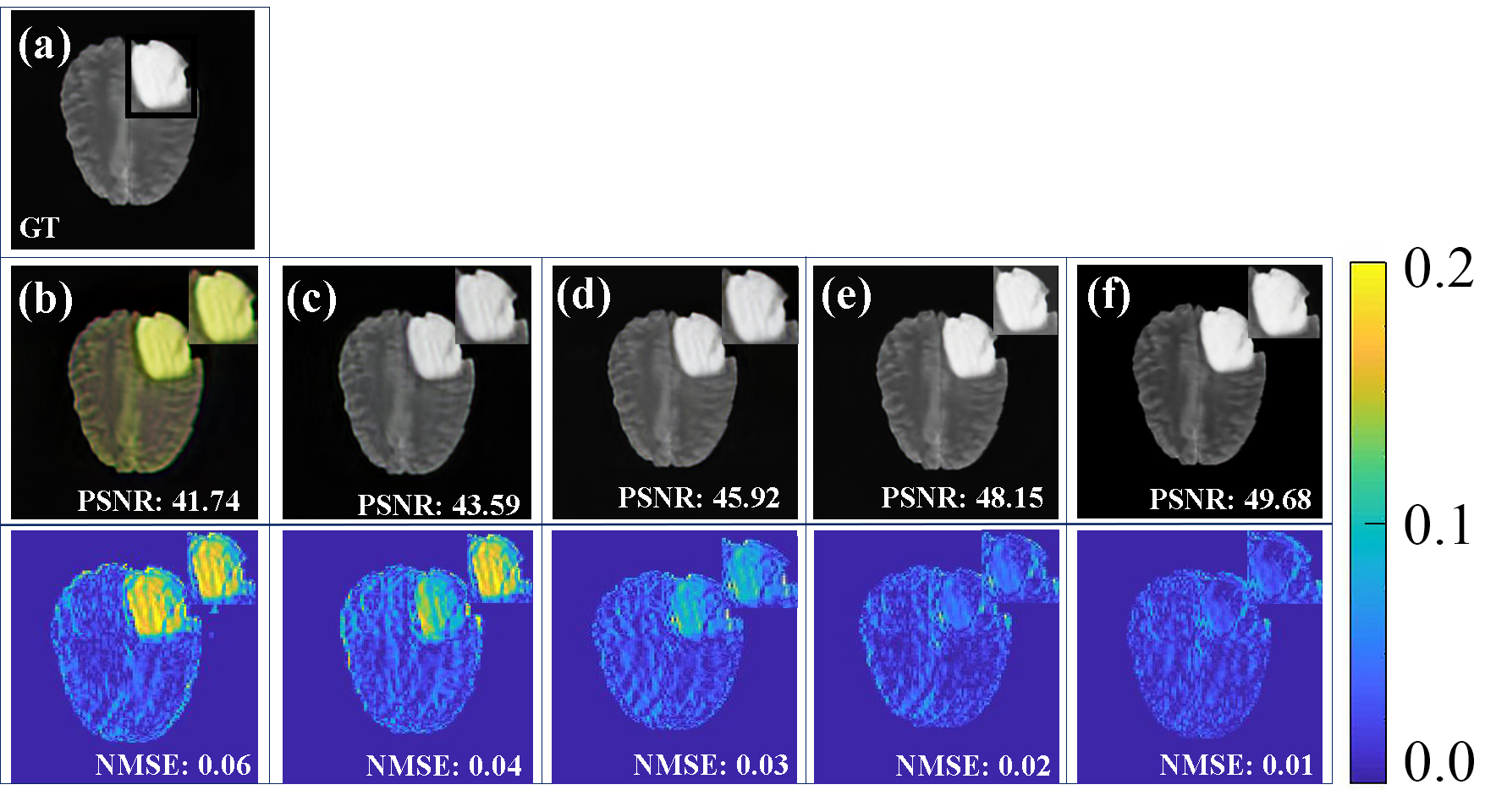}
    \caption{The reconstructed visual results of brain MRI for undersampling rate of 30$\%$ (acceleration factor of 3.3$\times$). (a) Ground truth, (b) DAGSP, (c) DAGSF, (d) DAGSF-GR, (e) REMGSF-GR, (f) RECGAN-GR.
    The difference image is placed below the corresponding reconstructed image. The specified region inside the black rectangular box is zoomed 4$\times$ and shown on the top right corner of each image.}
    \label{fig:Brain_figure}
\end{figure}
\begin{figure}[t]
    \centering
    \includegraphics[width=0.99998\linewidth]{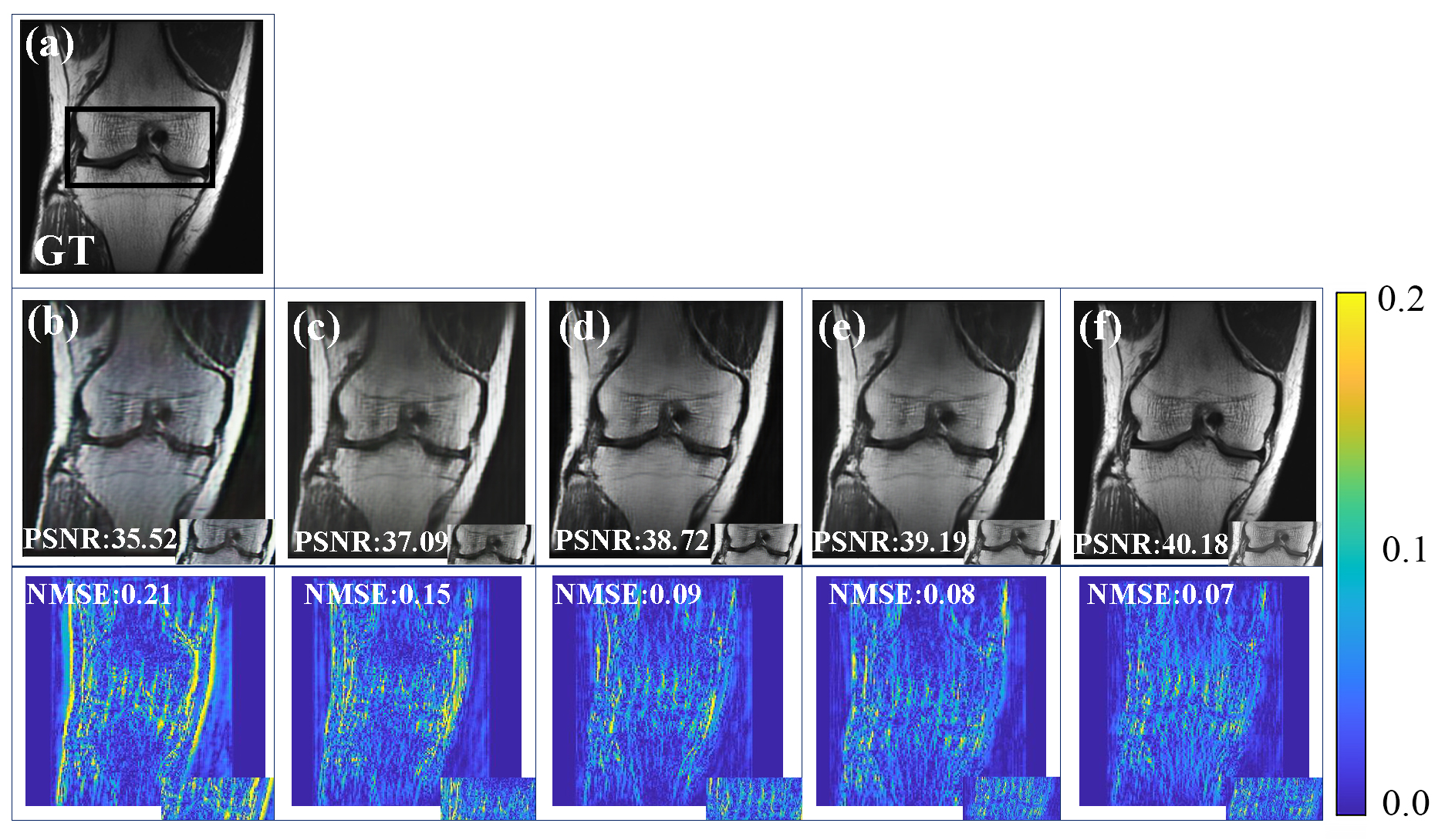}
    \caption{The reconstructed visual results of Knee MRI for undersampling rate of 30$\%$ (acceleration factor of 3.3$\times$). (a) Ground Truth, (b) DAGSP, (c) DAGSF, (d) DAGSF-GR, (e) REMGSF-GR, (f) RECGAN-GR. The difference image is placed below the corresponding reconstructed image. The specified region inside the black rectangular box is zoomed 4$\times$ and shown on the bottom right corner of each image. }
    \label{fig:Knee_figure}
\end{figure}
  The visual effects in Figs. \ref{fig:Brain_figure} and  \ref{fig:Knee_figure} are consistent with the quantitative assessment and the difference images show that RECGAN-GR can suppress artifacts and preserve more details than the aforesaid variants.
 To substantiate the generalizability of the proposed method, the parameters of the different variations are kept consistent for both the brain and the knee dataset.
\subsection{Comparison with Other Methods}\label{6:subsec2}
In this section, the effectiveness of the proposed RECGAN-GR is highlighted by comparing it with other CNN- and GAN-based methods using the brain and knee data.
\begin{table*}[h]\label{tab:brain_comp}
\caption{Performance comparison of different methods for Brain data with 1D Gaussian mask} 
\resizebox{18 cm}{1.5 cm}{
\centering 
\begin{tabular}{l ll ll ll ll ll} 
\hline 
Method  & \begin{tabular}{ c c  }
 \multicolumn{2}{c}{10\%}  \\ 
 \hline  PSNR & \hspace{0.8cm} SSIM  \\  
\end{tabular} &\begin{tabular}{ c c  }
 \multicolumn{2}{c}{20\%}  \\ 
 \hline  PSNR & \hspace{0.8cm} SSIM  \\  
\end{tabular}&\begin{tabular}{ c c  }
 \multicolumn{2}{c}{30\%}  \\ 
  \hline   PSNR & \hspace{0.8cm}  SSIM  \\  
\end{tabular} &\begin{tabular}{ c c  }
 \multicolumn{2}{c}{40\%}  \\ 
  \hline  PSNR & \hspace{0.8cm} SSIM  \\  
\end{tabular} &\begin{tabular}{ c c  }
 \multicolumn{2}{c}{50\%}  \\ 
  \hline  PSNR & \hspace{0.8cm}  SSIM  \\  
\end{tabular}
\\
\hline
RefineGAN  & \begin{tabular}{ c c }
 30.28 $\pm$ 4.19 & 0.880 
 \end{tabular}
   & \begin{tabular}{ c c   }
 32.72 $\pm$ 4.28 & 0.885 
 \end{tabular} & \begin{tabular}{ c c  }
 38.62 $\pm$ 6.05 & 0.923 
 \end{tabular} & \begin{tabular}{ c c  }
 41.30 $\pm$ 2.50 & 0.954 
 \end{tabular} & \begin{tabular}{ c c   }
 42.80 $\pm$ 4.65 & 0.956 
 \end{tabular}  \\
 DAGAN  & \begin{tabular}{ c c }
 32.52 $\pm$ 3.19 & 0.885 
 \end{tabular}
   & \begin{tabular}{ c c   }
 34.82 $\pm$ 6.56 & 0.894 
 \end{tabular} & \begin{tabular}{ c c  }
 41.57 $\pm$ 5.13 & 0.964 
 \end{tabular} & \begin{tabular}{ c c  }
 43.50 $\pm$ 5.41 & 0.973 
 \end{tabular} & \begin{tabular}{ c c   }
 44.80 $\pm$ 6.22 & 0.989 
 \end{tabular}  \\ 
DA-FWGAN & \begin{tabular}{ c c }
 33.02 $\pm$ 4.31 & 0.855
 \end{tabular}
   & \begin{tabular}{ c c   }
 34.23 $\pm$ 5.56 & 0.925 
 \end{tabular} & \begin{tabular}{ c c  }
 41.56 $\pm$ 5.27 & 0.984 
 \end{tabular} & \begin{tabular}{ c c  }
 43.56 $\pm$ 5.98 & 0.987 
 \end{tabular} & \begin{tabular}{ c c   }
 45.68 $\pm$ 6.31 & 0.990 
 \end{tabular}  \\
 W-Net-IIII & \begin{tabular}{ c c }
 35.30 $\pm$ 4.71 & 0.975
 \end{tabular}
   & \begin{tabular}{ c c   }
 34.68 $\pm$ 4.60 & 0.977 
 \end{tabular} & \begin{tabular}{ c c  }
 39.71 $\pm$ 6.00 & 0.980 
 \end{tabular} & \begin{tabular}{ c c  }
 42.08 $\pm$ 5.74 & 0.982 
 \end{tabular} & \begin{tabular}{ c c   }
 46.78 $\pm$ 6.23 & 0.984 
 \end{tabular}  \\
  DC-CNN & \begin{tabular}{ c c }
 35.07 $\pm$ 4.81 & 0.964
 \end{tabular}
   & \begin{tabular}{ c c   }
 35.48 $\pm$ 4.44 & 0.976 
 \end{tabular} & \begin{tabular}{ c c  }
 37.00 $\pm$ 4.48 & 0.981
 \end{tabular} & \begin{tabular}{ c c  }
 40.25 $\pm$ 5.74 & 0.984 
 \end{tabular} & \begin{tabular}{ c c   }
 48.28 $\pm$ 5.61 & 0.985 
 \end{tabular}  \\
 \hline
\textbf{RECGAN-GR}  & \begin{tabular}{ c c }
 \textbf{37.92 $\pm$ 3.41} & \textbf{0.972} 
 \end{tabular}
   & \begin{tabular}{ c c   }
\textbf{ 40.23 $\pm$ 3.85} & \textbf{0.982} 
 \end{tabular} & \begin{tabular}{ c c  }
 \textbf{46.20 $\pm$ 3.05} & \textbf{0.996} 
 \end{tabular} & \begin{tabular}{ c c  }
 \textbf{47.25 $\pm$ 4.35}  & \textbf{0.997}
 \end{tabular} & \begin{tabular}{ c c   }
 \textbf{49.87 $\pm$ 4.01} & \textbf{0.997} 
 \end{tabular}  \\ 
 \hline
\end{tabular}
}
\label{tab:brain_comp}
\end{table*}
 \begin{table*}[h]
\caption{Performance comparison of different methods for Knee data with 1D Gaussian mask} 

\resizebox{18cm}{1.5 cm}{
\centering 
\begin{tabular}{l ll ll ll ll ll} 
\hline 
Method  & \begin{tabular}{ c c   }
 \multicolumn{2}{c}{10\%}  \\ 
 \hline  PSNR & \hspace{0.8cm} SSIM  \\  
\end{tabular} &\begin{tabular}{ c c  }
 \multicolumn{2}{c}{20\%}  \\ 
 \hline  PSNR & \hspace{0.8cm} SSIM  \\  
\end{tabular}&\begin{tabular}{ c c  }
 \multicolumn{2}{c}{30\%}  \\ 
  \hline   PSNR & \hspace{0.8cm}  SSIM   \\  
\end{tabular} &\begin{tabular}{ c c   }
 \multicolumn{2}{c}{40\%}  \\ 
  \hline  PSNR & \hspace{0.8cm} SSIM \\  
\end{tabular} &\begin{tabular}{ c c }
 \multicolumn{2}{c}{50\%}  \\ 
  \hline  PSNR & \hspace{0.8cm}  SSIM  \\  
\end{tabular}
\\
\hline
 RefineGAN  & \begin{tabular}{ c c }
28.81 $\pm$ 3.32 & 0.913  
 \end{tabular}
   & \begin{tabular}{ c c   }
30.69 $\pm$ 4.28 & 0.922  
 \end{tabular} & \begin{tabular}{ c c }
 33.98 $\pm$ 4.43 & 0.930  
 \end{tabular} & \begin{tabular}{ c c  }
 34.72 $\pm$ 4.47 & 0.948  
 \end{tabular} & \begin{tabular}{ c c   }
 36.41 $\pm$ 5.02 & 0.949  
 \end{tabular}  \\
 DAGAN  & \begin{tabular}{ c c  }
 33.58 $\pm$ 4.31 & 0.895 
 \end{tabular}
   & \begin{tabular}{ c c  }
 34.01 $\pm$ 4.56 & 0.905 
 \end{tabular} & \begin{tabular}{ c c }
 35.29 $\pm$ 3.39 & 0.941 
 \end{tabular} & \begin{tabular}{ c c  }
 37.85 $\pm$ 2.91 & 0.960 
 \end{tabular} & \begin{tabular}{ c c  }
 38.89 $\pm$ 3.42 & 0.967  
 \end{tabular}  \\ 
DA-FWGAN & \begin{tabular}{ c c }
33.42 $\pm$ 2.09 & 0.901 
 \end{tabular}
   & \begin{tabular}{ c c   }
36.23 $\pm$ 3.96  & 0.912
 \end{tabular} & \begin{tabular}{ c c  }
37.26 $\pm$ 5.27 & 0.944 
 \end{tabular} & \begin{tabular}{ c c }
39.26 $\pm$ 3.22 & 0.965 
 \end{tabular} & \begin{tabular}{ c c  }
 39.55 $\pm$ 3.83 & 0.968 
 \end{tabular}  \\
  W-Net IIII & \begin{tabular}{ c c }
 24.83 $\pm$ 3.60 & 0.897
 \end{tabular}
   & \begin{tabular}{ c c    }
26.51 $\pm$ 4.35 & 0.920
 \end{tabular} & \begin{tabular}{ c c   }
 28.00 $\pm$ 4.50 & 0.922
 \end{tabular} & \begin{tabular}{ c c   }
  33.72 $\pm$ 3.95  & 0.933
 \end{tabular} & \begin{tabular}{ c c    }
43.76 $\pm$ 6.62 & 0.980
 \end{tabular}  \\ 
   DC-CNN & \begin{tabular}{ c c }
 31.59 $\pm$ 1.89 & 0.920 
 \end{tabular}
   & \begin{tabular}{ c c    }
 33.65 $\pm$ 1.32 & 0.933 
 \end{tabular} & \begin{tabular}{ c c   }
40.93 $\pm$ 3.32 & 0.984
 \end{tabular} & \begin{tabular}{ c c   }
 42.14 $\pm$ 3.78  & 0.944 
 \end{tabular} & \begin{tabular}{ c c    }
 44.40 $\pm$ 4.81 & 0.988 
 \end{tabular}  \\ 
\hline
\textbf{RECGAN-GR} & \begin{tabular}{ c c }
 \textbf{39.92 $\pm$ 2.43} & \textbf{0.971} 
 \end{tabular}
   & \begin{tabular}{ c c    }
 \textbf{40.05 $\pm$ 1.49} & \textbf{0.981} 
 \end{tabular} & \begin{tabular}{ c c   }
 \textbf{41.50 $\pm$ 1.67} & \textbf{0.992}
 \end{tabular} & \begin{tabular}{ c c   }
 \textbf{42.89 $\pm$ 1.78}  & \textbf{0.994} 
 \end{tabular} & \begin{tabular}{ c c    }
 \textbf{44.62 $\pm$ 1.87} & \textbf{0.997} 
 \end{tabular}  \\ 
 \hline
\end{tabular}
}
\label{tab:knee comp}
\end{table*} 
\begin{figure*}[!ht]
\centering
\includegraphics[width=0.9\linewidth]{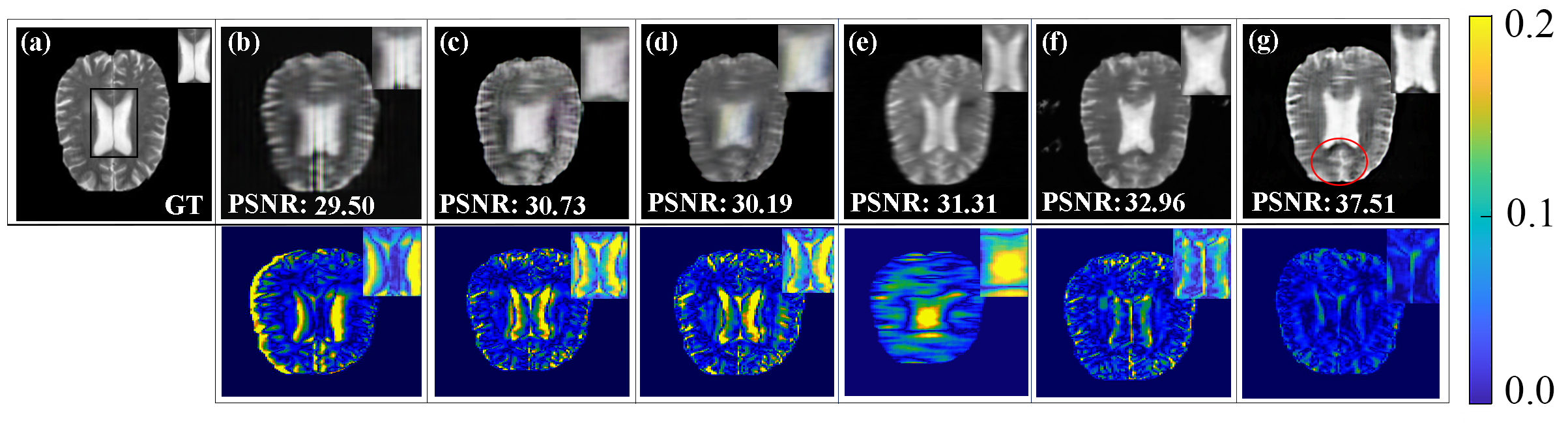}
\caption{Qualitative comparison of different  methods with the proposed RECGAN-GR for $10\%$ 1D Gaussian sampling. (a) Fully sampled reconstruction, (b) RefineGAN, (c) DAGAN, (d) DA-FWGAN, (e) DC-CNN, (f) W-Net IIII, (g) Proposed RECGAN-GR. The difference image is placed below the corresponding reconstructed image. The center region is zoomed 4$\times$ and shown on the top right corner of each image.}
\label{fig:brain_Figure_comp}
\end{figure*}
Table \ref{tab:brain_comp} summarizes the quantitative results of the five recent methods at different acquisition speeds (e.g., 10$\times$, 5$\times$, 3.3$\times$, 2.5$\times$, and 2$\times$) incorporating 1D Gaussian mask. Lower acquisition speeds corresponding to higher sampling rates lead to imperceptible visual detail among different reconstruction methods. On the contrary, higher acquisition speeds corresponding to lower sampling rates are more prone to aliasing. It is also evident from Table \ref{tab:brain_comp} that at 50$\%$ retained data the quantitative deviations in SNR among different methods are low. Significant improvement can be seen in the proposed RECGAN-GR for 20$\%$, 30$\%$ and $40\%$ sampling rates concerning other methods for brain MRI data. For 50$\%$ sampling, RECGAN-GR provides remarkable improvement over other methods except DC-CNN where the improvement is 3$\%$.  
Qualitative analysis for 10$\%$ sampling rate is shown in Fig. \ref{fig:brain_Figure_comp} which illustrates the loss of details due to higher acceleration. Early GAN-based methods suffer from loss of perceptual quality whereas DAGAN and DA-FWGAN improve significantly for the addition of VGG perceptual loss. Although, both methods struggle with the deprivation of details and blurring due to loss of information in the encoder and decoder path of the generator. Performances of DAGAN and DA-FWGAN are not much dissimilar as both the networks use the same loss functions though DA-FWGAN utilizes cascading three U-Net architectures to reconstruct the image. However, integrating more than one U-Net can contribute to notable information drop within the encoder-decoder paths. An increase in the variability for both brain and knee data shown in Tables \ref{tab:brain_comp} and \ref{tab:knee comp}, respectively demonstrate the problem of using sequential U-Nets for denoising. The error maps also support this observation and missing structural detail can be found at the center region for both DAGAN and DA-FWGAN. Although CNN-based methods embedded with the data consistency module perform better than the GAN-based methods, some motion blur due to aliasing can be seen. DC-CNN appears to have some serious motion artifacts which are distinctly visible in Fig. \ref{fig:brain_Figure_comp} and \ref{fig:Knee_Figure_comp}. However, W-Net IIII though performs significantly better than other methods but still suffers from loss of sharpness and the presence of little blurring. The proposed RECGAN-GR outperforms all the other methods performing better de-aliasing and preserving sharper details. The error map of the center region indicates the superior ability of detail recovery and, therefore, the magnified portion is less differentiable from the fully sampled reconstruction. Moreover, an interesting detail (sulcus) can be found in  Fig. \ref{fig:brain_Figure_comp} (g) (marked in red circle) which is only distinctly visible in the proposed RECGAN-GR whereas other methods lose more structural details of this small entity. 

\section{Discussion}\label{sec7}
The main purpose of this work is to emphasize the inclusion of parallel imaging-based loss function with compressed sensing MRI and deep learning. A conditional GAN-based network constrained with dual domain losses and parallel imaging based GRAPPA consistency loss is proposed here which combines the advantage of the parallel imaging method with CS-MRI for superior reconstruction. 

A novel generator architecture RemU-Net is also introduced to address the problem of information loss from encoder-decoder along with a \textit{k}-space correction block as refinement learning. The comparative studies explicitly indicate better reconstruction both statistically and visually using the multi-loss function scheme. The idea of the weighted phase loss function and parallel imaging-based loss functions can also be integrated with other methods such as DC-CNN and W-Net-IIII to improve the reconstruction at a lower sampling rate and achieve faster reconstruction. The idea of RemU-Net can be modified for other GAN networks such as Cycle-GAN and Style-GAN type networks \cite{Sandfort2019,Karras2020}. The \textit{k}-space correction block can be efficiently altered into data consistency blocks for further variation of the RECGAN-GR network. 
 \begin{figure}[t]
\centering
\includegraphics[width=0.9\linewidth]{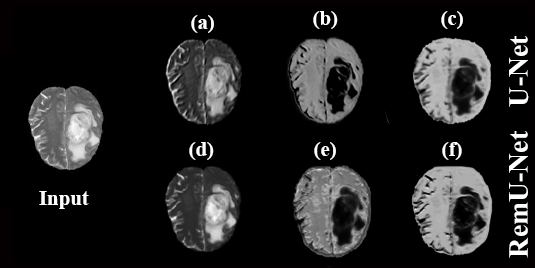}
\caption{Visualization of the feature maps of the RemU-Net and U-Net. The first row represents three-channel features a, b, and c respectively at the last convolution layer of U-Net. The second row represents three-channel features d, e, and f respectively at the last convolution layer of RemU-Net. Some images have their contrast reversed for improved visualization.}
\label{fig:U vs REM}
\end{figure}
 Medical image reconstruction demands a higher level of accuracy but skip connections cannot single-handedly conserve the deep, coarse-grained feature from the encoder to decoder network. For the underlying problem, the input image is aliased and some latent features cannot be recovered from blurry images even with the skip connections. The remnant block extracts the salient features from the image \cite{Rafi2020} thereby, preserving these features in the proposed RemU-Net generator. As shown in Fig. \ref{fig:U vs REM}, the RemU-Net preserves more details (e) and sharpness (d, f) than the U-Net architecture used in DAGAN \cite{Yang2018}.

The major advantage of the proposed method which is distinct from other deep learning-based compressed sensing imaging methods is that it can be implemented with parallel imaging such as GRAPPA. As parallel imaging is the default option for most MRI scanning machines, the proposed method can be integrated explicitly with practical MRI scanning protocols. Reconstruction quality mostly depends on the sampling process and compressed sensing allows multiple sampling techniques. Most of the CS-MRI methods provide improved reconstruction when sampling pattern adapts with the probability of the \textit{k}-space points of variable random density such as 2D Gaussian, 2D Poisson functions \cite{Vellagoundar2015}. The implementation of random 1D sampling masks requires the changing of amplitudes of the phase encoding gradients randomly whereas 2D sampling masks require simultaneous change of the phase encoding and frequency encoding gradients randomly which is not feasible to conventional MRI scanners. However, by varying both the amplitude and direction of the phase encoding gradients, it is possible to obtain pseudo-random sampling masks in practical scanners \cite{HaifengWang2009}. 
\begin{table}[h]
\caption{Performance results of different methods on Brain data for 20\% 2D Random sampling patterns} 
\resizebox{9 cm}{1.3 cm}{
\centering 
\begin{tabular}{l ll ll} 
\hline 
Method  & \begin{tabular}{ c c  }
 \multicolumn{2}{c}{2D Gaussian}  \\ 
 \hline  PSNR & \hspace{0.8cm} SSIM  \\  
\end{tabular} &\begin{tabular}{ c c  }
 \multicolumn{2}{c}{2D Poisson}  \\ 
 \hline  PSNR & \hspace{0.8cm} SSIM  \\  
\end{tabular} 
\\
\hline
 DAGAN  & \begin{tabular}{ c c }
37.79 $\pm$ 6.02 & 0.902 
 \end{tabular}& \begin{tabular}{ c c  }
 42.12 $\pm$ 6.39 & 0.996 
 \end{tabular} 
\\
DA-FWGAN & \begin{tabular}{ c c }
39.02 $\pm$ 4.12 & 0.972
 \end{tabular}
   & \begin{tabular}{ c c   }
41.33 $\pm$ 4.00 & 0.977
 \end{tabular} 
\\
 W-Net-IIII & \begin{tabular}{ c c }
 42.38 $\pm$ 6.34 & 0.994
 \end{tabular}
   & \begin{tabular}{ c c   }
45.23 $\pm$ 3.56 & 0.997 
 \end{tabular} 
 \\
  DC-CNN & \begin{tabular}{ c c }
 35.23 $\pm$ 3.25 & 0.949
 \end{tabular}
   & \begin{tabular}{ c c   }
 38.56 $\pm$ 3.56 & 0.955 
 \end{tabular} 
\\
 \hline
\textbf{RECGAN-GR}  & \begin{tabular}{ c c }
 \textbf{46.62 $\pm$ 3.41} & \textbf{0.997} 
 \end{tabular}
   & \begin{tabular}{ c c   }
\textbf{48.19 $\pm$ 4.89} & \textbf{0.998}
 \end{tabular}  \\ 
 \hline
\end{tabular}
}
\label{tab:PPer1}
\end{table}
Table \ref{tab:PPer1} presents a statistical analysis of 20$\%$ sampling (5$\times$ acquisition) for 2D Gaussian and 2D Poisson sampling for the test brain MR images. The overall metrics are better than 1D Gaussian sampling which is consistent with the previous studies reported on these sampling schemes. It is also evident that our proposed RECGAN-GR outperforms all other methods in case of the 2D Gaussian and 2D Poisson sampling patterns making it a vigorous tool to adapt in various kinds of sampling schemes.
 \begin{figure}[htp]
\centering
\includegraphics[width=0.9998 \linewidth]{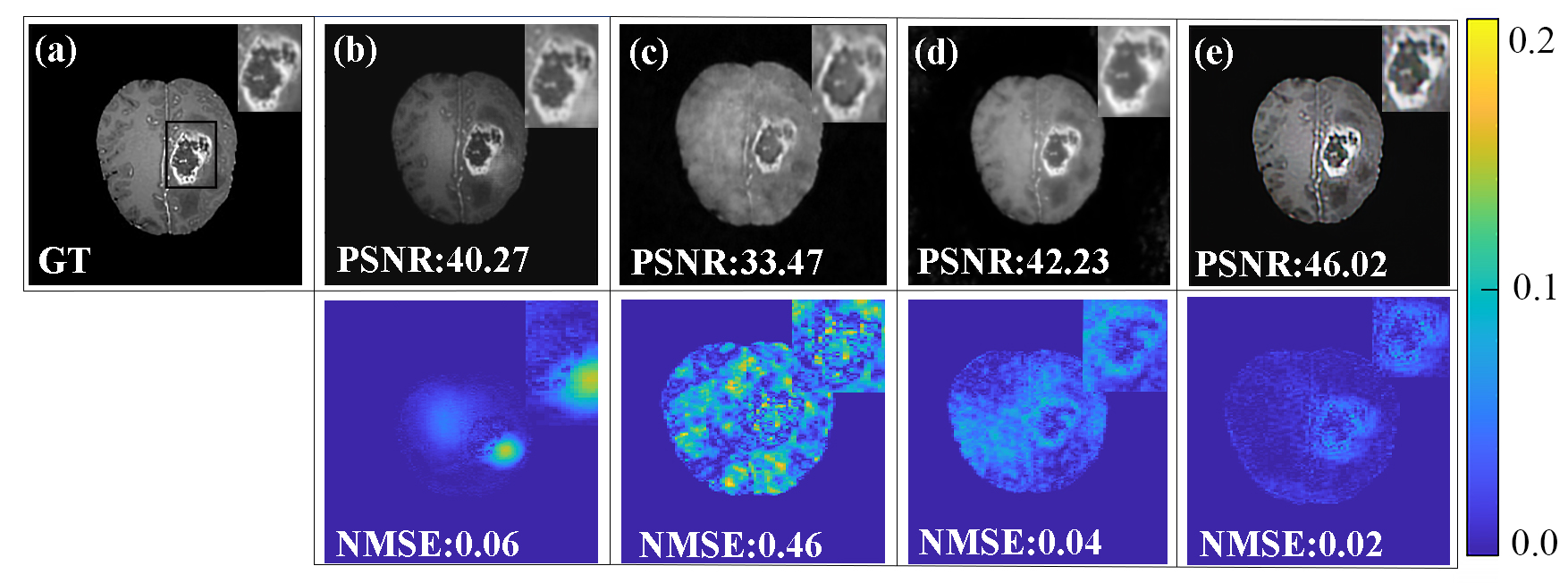}
\caption{Qualitative comparison of different methods with proposed RECGAN-GR with 20 $\%$ random 2D Gaussian sampling (a) Fully sampled reconstruction, (b) DAGAN, (c) DC-CNN, (d) W-Net IIII, (e) Proposed RECGAN-GR. The infectious region is zoomed 4$\times$ and shown on the top right corner of each image.}
\label{fig:brain_Figure_2D}
\end{figure}
Another important aspect of MRI is flawless reconstruction of anatomical or pathological anomaly. Fig. \ref{fig:brain_Figure_2D} represents a qualitative visualization of brain region reconstruction from 20$\%$ random 2D Gaussian sampling mask. The error maps of 4$\times$ zoomed region distinctly indicates the superior performance of the proposed RECGAN-GR ensuring anomaly reconstruction at 5$\times$ acquistion speed.
In case of processing time, a methodical
comparison with other methods is not possible due to the fact that the different methods were
implemented in different setups. In addition, fine-tuning of the hyper-parameters of different methods heavily depends on the experimental setups. 
 \begin{table}[b]
 \centering
 \caption{Comparison of parameters and reconstruction speed of different methods} 
 \resizebox{8 cm}{1.3 cm} {
 \begin{tabular}{ |c|c|c|c| } 
 \hline
 \textbf{Methods} & \textbf{Number of parameters} & \textbf{Reconstruction speed} \\
 \hline
DA-FWGAN & 21,78,15,930 & 0.315s \\ 
 \hline
  RefineGAN & 14,69,49,095 & 0.294s  \\ 
  \hline
 DAGAN & 7,60,82,260 & 0.217s \\ 
  \hline
\textbf{RECGAN-GR} & \textbf{7,83,75,264} & \textbf{0.192s} \\ 
 \hline
  DC-CNN  & 5,04,876 & 0.03s \\ 
  \hline
  W-Net IIII & 1,13,31,299 & 0.01s\\ 
 \hline
 \end{tabular}
 }
 \label{tab:speedc}
\end{table}
However, the number of parameters and time needed to
reconstruct a single image in the setup mentioned in section \ref{5:subsec2} is summarized in Table \ref{tab:speedc} for all the methods to ensure a fair comparison.
Although it is apparent from the Table \ref{tab:speedc} that the proposed RECGAN-GR surpasses all the GAN-based methods in terms of reconstruction speed it still lags behind the CNN-based methods. The number of parameters indicate the CNN-based methods are lightweight compared to the GAN-based methods. The GANs train two networks simultaneously (the generator and the discriminator) compared to a single CNN which leads to longer reconstruction time. However, the time difference is very small (e.g., 0.162s) compared to DC-CNN which makes RECGAN-GR, a worthy candidate for fast MRI acquisition with better reconstruction quality.

 \begin{figure}[t]
\centering
\includegraphics[width=0.99998 \linewidth]{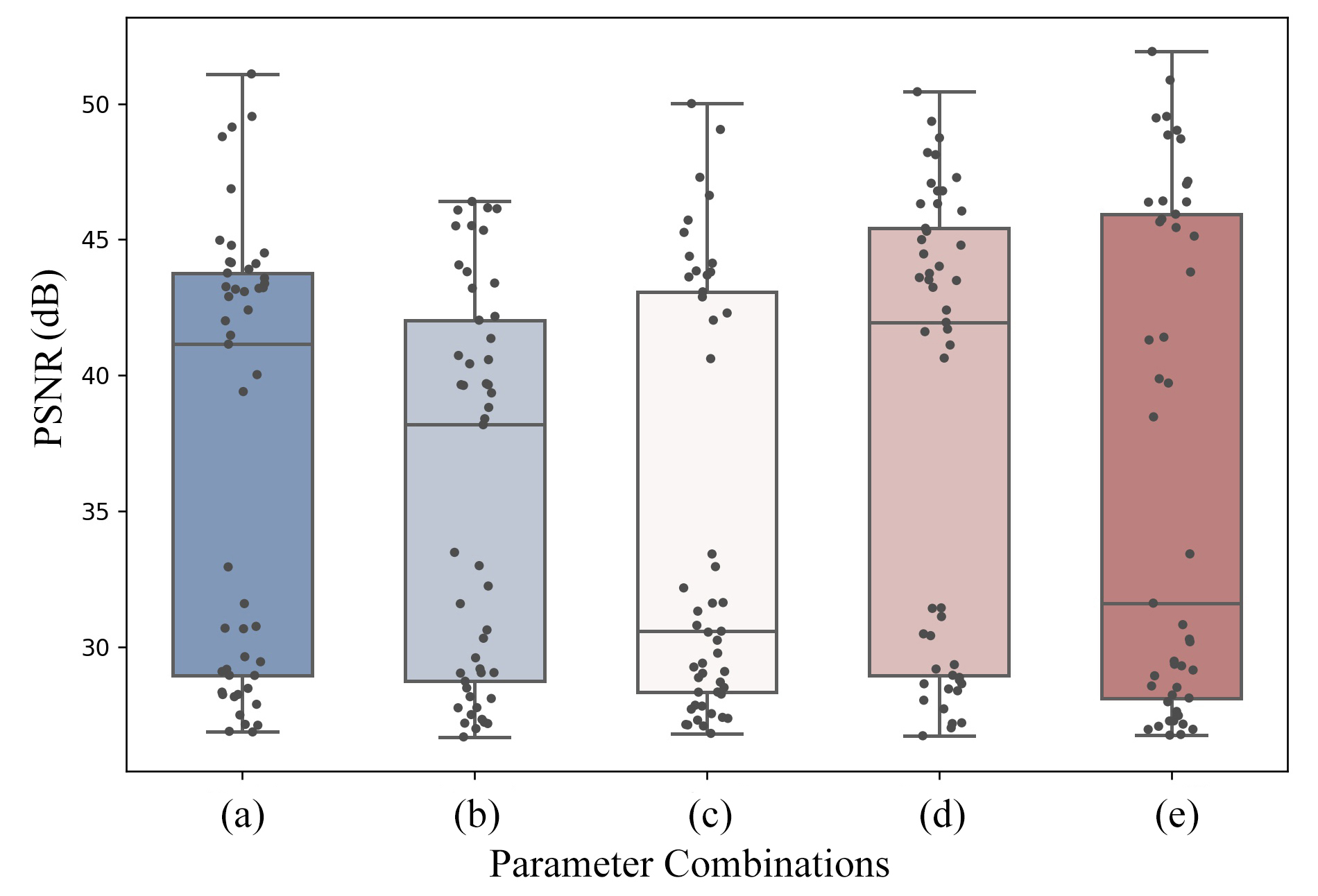}
\caption{The impact of hyperparameters tuning on the reconstruction quality,
utilizing a Cartesian mask with R = 2, A = 9, and RECGAN-GR to reconstruct the multi-coil test data. (a) $\alpha = 12, \beta = 0.08, \gamma = 0.06, \delta = 0.02, \zeta = 0.00030$, (b) $\alpha = 14, \beta = 0.09, \gamma = 0.04, \delta = 0.009, \zeta = 0.00020$, (c) $\alpha = 15, \beta = 0.1, \gamma = 0.05, \delta = 0.01, \zeta = 0.00025$, (d) $\alpha = 16, \beta = 0.2, \gamma = 0.04, \delta = 0.008, \zeta = 0.00015$, (e) $\alpha = 17, \beta = 0.3, \gamma = 0.02, \delta = 0.006, \zeta = 0.00010.$ }
\label{fig:param}
\end{figure}

 \begin{figure}[t]
\centering
\includegraphics[width=0.99998 \linewidth]{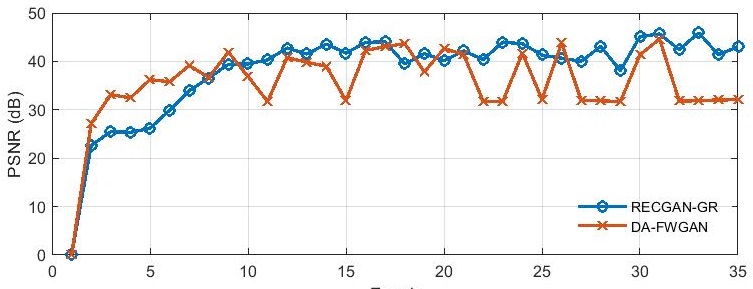}
\caption{Convergence analysis of the proposed RECGAN-GR with respect to DA-FWGAN. Here, random 2D Gaussian sampling mask is used for a test brain MR image.}
\label{fig:speed}
\end{figure}

Setting up the parameters of loss functions, i.e., $\alpha$, $\beta$, $\gamma$, $\delta$, and $\zeta$ is crucial for efficient training time and stability of GAN. To come up with the parameters mentioned in \ref{5:subsec2}, extensive simulations are performed for RECGAN-GR with Cartesian sampling schemes on the multi-coil dataset. We tried five different combinations of parameters which are summarized as box plot in Fig. \ref{fig:param}. The plot represents the range of PSNR in different images of the test set. It is evident from Fig. \ref{fig:param} that our proposed RECGAN-GR is not susceptible to hyperparameters tuning. However, the set of parameters (b) shows inferior results whereas the set of parameters (d) presents the maximum median value with most of the samples being accumulated on the high PSNR side. Therefore, the set of parameters (d) is chosen for the simulation study of RECGAN-GR. Furthermore, parameter $\kappa$ is kept constant at $10^{-3}$, otherwise, it severely produces checkerboard artifacts and hinders the stable convergence of RECGAN-GR.

The convergence analysis is another important aspect of training the GAN network. In this paper, the proposed \textit{k}-space correction block is forcing the generator to reconstruct the missing lines to avoid the mixing of noisy data with the original ones. A comparison of refinement learning used in the DAGAN and DA-FWGAN \cite{Yang2018,Jiang2019} with the proposed RECGAN-GR method is illustrated in Fig. \ref{fig:speed}. Here, DA-FWGAN is shown instead of DAGAN because DA-FWGAN is a fine-tuned version of DAGAN. The proposed RECGAN-GR minimizes more loss functions compared to DAGAN or DA-FWGAN which leads to slower convergence below the first 10 epochs. However, after 15 epochs the proposed RECGAN-GR stabilizes with better reconstruction quality compared to DA-FWGAN. In particular, the proposed method maintains consistency due to the proper utilization of the \textit{k}-space correction block which prevents the mixture of noisy \textit{k}-space lines with the original raw lines. DA-FWGAN fails to reach an optimum solution as more and more noise mixing occurs.

Despite the fact that, the proposed method takes longer reconstruction than the CNN methods, it delivers better reconstruction quality and faster reconstruction than the GAN-based methods that exist in the literature. The superior reconstruction quality and consistency make RECGAN-GR an eligible candidate for MR image reconstruction. The fact that most deep learning based algorithms lack robustness is another one of their significant issues \cite{Akakaya2018}. We tested our proposed RECGAN-GR, not only with different datasets but also with different sampling schemes, e.g., random and Cartesian, making it a robust deep learning based method for fast MRI reconstruction.
\section{Conclusion}\label{sec8}
In this paper, a novel conditional GAN is introduced for MRI reconstruction where RemU-Net generator, in consistent with the U-Net architecture is incorporated for better preservation of semantic features which leads to the superior reconstruction of anatomical details. The network is constrained to dual-domain weighted loss functions in parallel with GRAPPA consistency loss. The \textit{k}-space correction block proposed in our scheme preserves the originally acquired lines and prevents further distortion or spread of noise and thereby ensuring consistent reconstruction and fast convergence. RECGAN-GR
can efficiently remove artifacts while preserving more details at
faster acquisition speeds (5$\times$, 10$\times$) with different
sampling strategies (1D Gaussian, random 2D Gaussian, and random 2D Poisson
sampling). Compared to several
recent methods (RefineGAN, DAGAN, DA-FWGAN,
DC-CNN and W-NET IIII), the proposed model can successfully reconstruct the MRI images with better preservation of detail from highly under-sampled \textit{k}-space data. The components of the proposed architecture can be blended into traditional GAN networks and have a significant role not only in MRI but also in other image reconstruction applications such as CT, SPECT, and PET  image reconstruction.
\section{Declaration of Competing Interest}
\label{CI}
The authors state that they have no competing financial interests or personal relationships that may have influenced the work presented in this study.
\bibliographystyle{ieeetr}
\bibliography{main.bib} 
\end{document}